\documentclass[aps,pre,epsf,superscriptaddress,amsmath,amssymb,amsfonts,twocolumn,showpacs]{revtex4-1}
\usepackage{amsmath,amssymb,amsfonts,bm,comment}
\usepackage{graphics,float,tikz}
\usepackage{appendix}
\usetikzlibrary{arrows.meta,arrows}
\usepackage[caption=false]{subfig}
\usepackage[colorlinks=true,
	linkcolor=blue,
	filecolor=magenta,      
	urlcolor=blue,
	citecolor=blue]{hyperref}
\usepackage{physics,xparse,mathtools,isotope}
\usepackage[table-parse-only]{siunitx}
\usepackage{multirow,array,relsize,lipsum,longtable,tabularx}
\usepackage{dsfont}

\newcommand{\im}{\mathrm{i}}

\usepackage{amsthm}

\newtheorem{remark}{Remark}



\begin{document}
	\title{Synchronized Bell protocol for detecting non-locality between modes of light}

	\author{Madhura Ghosh Dastidar}
	\affiliation{Department of Physics, Indian Institute of Technology Madras, Chennai 600036, Tamil Nadu, India}
	
	\author{Gniewomir Sarbicki}
	\affiliation{Institute of Physics, Faculty of Physics, Astronomy and Informatics, Nicolaus Copernicus University, Grudzi\c adzka 5/7, 87-100 Toru\'n, Poland}
	
	\author{Vidya Praveen Bhallamudi}
	\affiliation{Quantum Center of Excellence for Diamond and Emerging Materials (QuCenDiEM) Group, Departments of Physics and Electrical Engineering, Indian Institute of Technology Madras, Chennai 600036, India.}
	
	\date{\today}
	
	\begin{abstract}
		
    In the following paper, we discuss a possible detection of non-locality in two-mode light states in the Bell protocol, where the local observables are constructed using displacement operators, implemented by Mach-Zender Interferometers fed by strong coherent states. We report numerical results showing that maximizing the Braunstein-Caves Chained Bell (BCCB) inequalities requires equal phases of displacements. On the other hand, we prove that non-locality cannot be detected if the phases of displacements are unknown. Hence, the Bell experiment has to be equipped with a synchronization mechanism. We discuss such a mechanism and its consequences.
    
	\end{abstract}
	
	\maketitle
	
	\section{Introduction}
    
    Entangled quantum systems have grown in importance for technological as well as fundamental scientific applications. The advantage of quantum non-locality has been proved in various fields such as quantum communication~\cite{yuan2010entangled,sangouard2010quantum}, metrology~\cite{giovannetti2006quantum,sabines2017sub} and computation~\cite{o2007optical,briegel2009measurement}. Entangled modes of light typically, are useful in photonic quantum metrology schemes~\cite{polino2020photonic}, where the purpose is to achieve the quantum limit of measurement~\cite{wiseman2010quantum}. These states of light are multiphotonic, i.e., combinations of superpositions of Fock states. Thus, the experimental verification of entanglement in such states requires many measurements with complex experimental setups. For example, recent works~\cite{israel2019entangled,li2021experimental} show that experimental verification of entanglement in certain important classes of two-mode entangled states require multiple single photon detectors or photon-counting electron-multiplying charge-coupled-device (EMCCD) camera. In this approach, the density matrix is reconstructed in the process of full-state tomography which requires restriction to an effective Fock space of dimension $n$ and the number of observables to be measured grows fast with $n$. 

    As an alternative to performing such intricate experimental schemes, one can perform a Bell-CHSH~\cite{bell1964einstein,clauser1969proposed} experiment as proposed in~\cite{dastidar2022detecting} for entanglement detection in two-mode light states. This work describes using Mach-Zehnder Interferometers (MZI) fed with a strong coherent state at one input port and having a photodetector at one output port. The photodetector can measure zero or non-zero intensities of the incoming pulse. This experimental unit (MZI + coherent state + photodetector) is possessed by each of two parties. The above is relatively simpler compared to the existing schemes for verification of entanglement in two modes of light.
 
    A CHSH inequality is defined for two parties with two measurement settings ($n=2$) per party. The Braunstein-Caves chained Bell (BCCB) inequalities~\cite{braunstein1990wringing} generalise the CHSH inequality to $n$ measurement settings per party. A particular expression for the quantum bound of BCCB inequalities 
    has been reported in~\cite{wehner2006tsirelson}. It is also shown there that the difference between quantum and classical bound grows with $n$ for $n>2$.
    Thus, in an experiment, the violation of the classical bound by a two-mode entangled state can be resolved better with $n>2$. 
    
    In the following paper we check, whether the CHSH inequality in the mentioned experimental scheme can be improved by using BCCB inequality when the parties again use observables implemented by MZI + coherent state + photodetector.

     In this paper, we intend to check if such a generalization can be extended to the proposed setup in~\cite{dastidar2022detecting}. We observe that the Mach-Zehnder interferometric setup involved in entanglement detection requires phase synchronization of the two inputs to the interferometer. We report that without a constant phase difference between the two inputs, the measurement observables get restricted to the classical regime. Thus, entanglement detection is only possible when there is a known and fixed phase difference between the two inputs of the MZI. We also discuss the two-mode light states for which this setup is best for the experimental detection of entanglement. 
     
     Further, the entanglement detection in the scheme should be also analysed under restriction to experimentally accessible classes of entangled two-mode light states. We check whether the proposed experimental scheme detects entanglement for certain important states of light useful for quantum metrology, namely, entangled coherent states (ECS)~\cite{sanders2012review} and two-mode squeezed vacuum (TMSV)~\cite{hiroshima2001decoherence}.

    The paper is organized as follows: Sec.~\ref{BCCBi} describes the formulation of the BCCBI inequality for $n$ measurement settings per party for our proposed experimental setting. In Sec.~\ref{Results}, we report our numerical results of maximal violation obtained by the $n$-MZI settings 
    and comment on the phase synchronization issues. We also give a brief description of the states that correspond to this maximal violation. In Sec.~\ref{Optimal Violation for Experimentally Achievable States}, we consider entanglement detection for two important classes of light: entangled coherent states and two-mode squeezed vacuum, and discuss the values of parameters maximizing the violation obtained by $n$-MZI settings. We summarise our observations in Sec.~\ref{Conclusion}.
	
	\section{The Braunstein-Caves chained Bell (BCCB) inequality} \label{BCCBi}
	
	In general, for $n$ dichotomic observables (of output values $\pm 1$) per party, the following inequality holds under the assumption of the existence of underlying probability space (local hidden variable model):
	\begin{align}\label{genCHSH}
    |\mathbb{E}(\sum_{i=1}^n X_i \otimes Y_i + \sum_{i=1}^{n-1} X_{i+1}\otimes Y_i - X_1\otimes Y_n)| \le 2n-2.
    \end{align}
    where $\{X_1,..., X_n\}$ and $\{Y_1,...,Y_n\}$ are
	dichotomic observables employed by Lab X and Y, respectively, corresponding to their $n$ independent measurement settings.
    The above inequality is known as the Braunstein-Caves chained Bell (BCCB) inequality and the maximum of the LHS over all quantum states is $2n\cos(\frac{\pi}{2n})$~\cite{wehner2006tsirelson}.
    The bound is saturated for the qubit singlet state $(\ket{00}+\ket{11})/\sqrt{2}$ and observables:
    \begin{align}
        X_i = \cos(\alpha_i) \sigma_x + \sin(\alpha_i) \sigma_y = 
        \left[ \begin{array}{cc} 0 & e^{-i\alpha_i} \\ e^{i\alpha_i} & 0
        \end{array} \right]
        \label{XPauli}
        \\
        Y_i = \cos(\beta_i) \sigma_x + \sin(\beta_i) \sigma_y = 
        \left[ \begin{array}{cc} 0 & e^{-i\beta_i} \\ e^{i\beta_i} & 0
        \end{array} \right]
        \label{YPauli}
    \end{align}
    where $\alpha_k = k\pi / n$, $\beta_k = -k\pi / n$.
    
    For $n=2$ the BCCB inequality becomes the famous CHSH inequality.
    
\begin{figure*}[ht]
		\centering
		\includegraphics[width=0.9\linewidth]{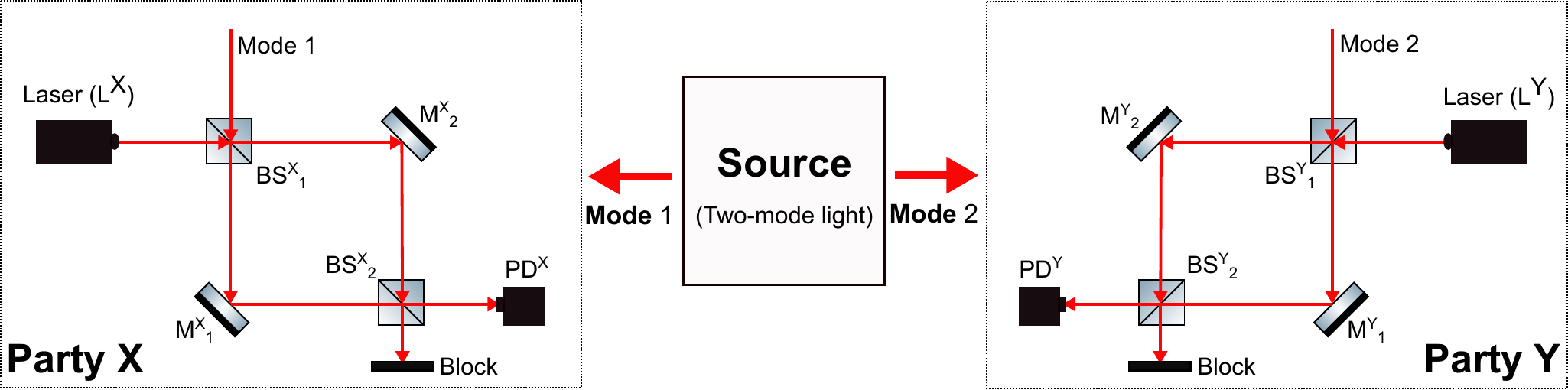}
		\caption{\textbf{Schematic of experimental arrangement for entanglement detection:} The proposed setting with a source producing a two-mode entangled state of light and 2 laboratories (X and Y) involved in a Bell-type experiment. Each party uses a Mach-Zehnder interferometer (MZI), comprising 50:50 beam-splitters BS$^j_i$ and mirrors M$^j_i$ ($i = \text{X or Y}, j = 1\text{ or }2$). Each MZI is fed with a coherent state $\ketbra{\alpha}_i$ at one input and is terminated by a photodetector PD$_i$ ($i = \text{X or Y}$) which measures zero or non-zero intensities. The two paths in the MZI have a relative phase difference $\phi_i$ ($i = \text{X or Y}$).}
		\label{MZI_setup}
    \end{figure*}

    Let us assume, that each party performs intensity-based measurements on its mode using a photodetector at the output of Mach-Zehnder interferometer (MZI), where its first input is fed by the possessed mode and the second by a strong coherent state of light (Figure \ref{MZI_setup}). Such interferometer setting implements a displacement operator $\hat{D}(\alpha)$ on the input mode and, together with the photodetector, a projective measurement: $\{ \ket{\alpha}\bra{\alpha}, I - \ket{\alpha}\bra{\alpha} \}$. Prescribing output values $\pm 1$, we obtain a hermitian observable:
    \begin{equation}\label{Abeta}
	    A(\alpha) = \mathbb{I} - 2\ketbra{\alpha}{\alpha}.
	\end{equation}
    The $n$ measurement settings on each side correspond to $n$ displacements.
    
    Let the measurement settings or displacements implemented by MZIs in Lab X and Y be $\{\beta_1,..., \beta_n\}$ and $\{\gamma_1,..., \gamma_n\}$, respectively. The corresponding observables
	are $\{A(\beta_1),..., A(\beta_n)\}$ and $\{A(\gamma_1),..., A(\gamma_n)\}$.

    Therefore, the observables in [Eq.~\ref{genCHSH}] are:
	
	\begin{align}\label{X_and_Y}
	    X_i &= A(\beta_i) = \mathbb{I} - 2\ketbra{\beta_i}{\beta_i} \nonumber \\
	    Y_i &= A(\gamma_i) = \mathbb{I} - 2\ketbra{\gamma_i}{\gamma_i}
	\end{align}

    Thus, we can write the LHS of BCCB inequality ($S$) for $n$-MZI settings from [Eq.~\ref{genCHSH}] as:
    \begin{equation}\label{S_beta_gamma}
	    S = \sum_{i=1}^n A(\beta_i) \otimes A(\gamma_i) + \sum_{i=1}^{n-1} A(\beta_{i+1}) \otimes A(\gamma_i) - A(\beta_1) \otimes A(\gamma_n)
	\end{equation}
	
	In \cite{dastidar2022detecting}, it has been proven, that for $n=2$, the maximal violation of the CHSH inequality can be achieved by an appropriate choice of displacements in both (MZI+photodetector) settings possessed by Lab X and Lab Y. Now, to detect entanglement by a larger ($n>2$) number of settings, the classical bound ($2n-2$) must be violated, i.e., $(2n-2)<\mathbb{E}(S)_{max}\le2n\cos{(\pi/2n)}$.
	Further, we check the MZI settings in both labs maximizing the violation and how close to the maximal violation $2n\cos{(\pi/2n)} - (2n-2)$ can it be.

	\section{Results}\label{Results}
    In this section, we discuss a number of optimization results we have obtained analysing the BCCB inequality with observables originating from MZI setups and for various families of experimentally accessible states.
    We have obtained the results numerically and the Appendices~\ref{A: Orthonormal Basis} and~\ref{A: n-MZI} describe the details of our codes.
    
     \begin{figure*}[ht]
		\centering
		\includegraphics[width=\linewidth]{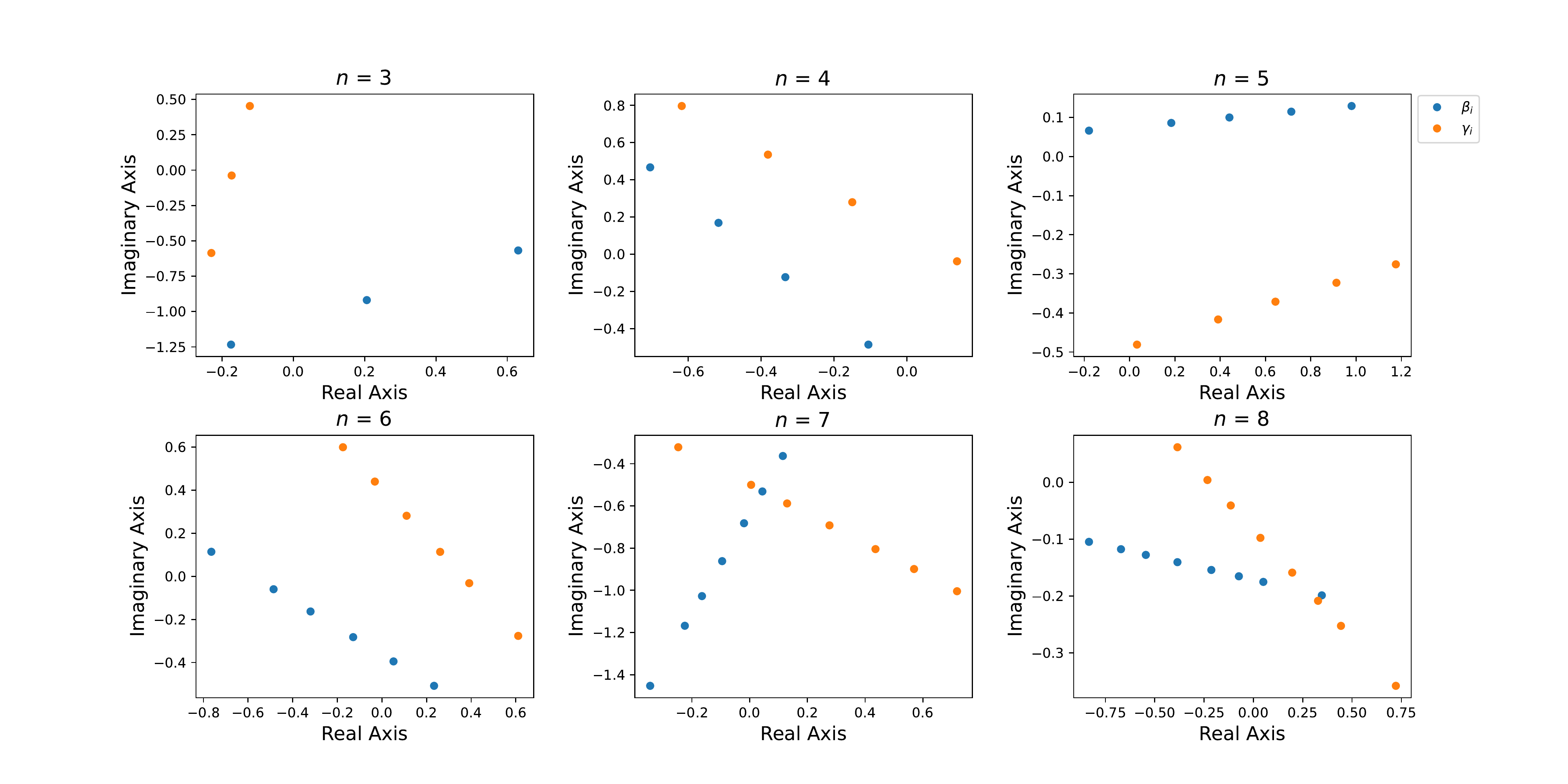}
		\caption{\textbf{Complex-plane representation of $\{\beta_i\}$ and $\{\gamma_i\}$:}
		Numerically generated plots for the optimized $\{\beta_i\}$ and $\{\gamma_i\}$ ($i = 1\text{ to }n, n\in [3,5]$) for which the corresponding maximal violation of BCCB inequality [Eq.~\ref{genCHSH}] is achieved. A co-linear trend is observed in the complex-plane representation of these $\{\beta_i\}$ and $\{\gamma_i\}$ and thus, each may be written as approximate arithmetic sequences.
		}
		\label{complex_plane_beta_gamma}
\end{figure*}
    
    \subsection{Maximal Eigenvalues of BCCB matrix}\label{Maximal Eigenvalues of CHSH matrix}

    As discussed earlier,
    the BCCB inequality is maximally violated in a pure state represented by an eigenvector of $S$ (\ref{S_beta_gamma}) related to its maximal eigenvalue. The maximum possible violation is equal to $2n\cos{(\frac{\pi}{2n})} - (2n-2)$, and in particular for $n=2$, we have obtained $2\sqrt{2} - 2$ $-$ the maximal violation for standard CHSH inequality. 
    First, we perform optimization for $n = 3,\dots,8$ with respect to the parameters $\{\beta_i\}, \{\gamma_i\}$, for $i \in \{1, \dots, n\}$. 
    We minimize the probability of getting stuck in a local maximum by repeating the procedure multiple times, with a number of randomly chosen starting points.
    At this stage, for $n>10$ the method typically gets stuck in a local minimum and shows no violation. 
    
    The optimal sequences of $\{\beta_i\}$ and $\{\gamma_i\}$ are shown in Fig.~\ref{complex_plane_beta_gamma}.
    We observe that the sequences $\{\beta_i\}$ and $\{\gamma_i\}$ each behave co-linearly on the complex plane. The common phase of $\beta_i$ can be made zero by applying a local unitary transformation. Similarly, the first displacement (measurement setting) can be made zero by applying a displacement operator, which is a local unitary transformation as well.
    The same applies to co-linear complex numbers $\gamma_i$.
    Hence both sequences are real and start from $0$.
    In this way, we reduced the number of optimization parameters from $4n$ to $2n-2$. 
    
    Moreover, in each sequence, we observe almost equal spacing between displacements except for the first/last one, being significantly bigger [see Fig.~\ref{complex_plane_beta_gamma}].
    We confirm this observation in the optimization over a reduced number of parameters, obtaining almost perfect matching with a two-parameter optimization, where $\beta_2 - \beta_1 = \gamma_n - \gamma_{n-1} = \Delta'$, $\beta_{i+1}-\beta_i = \Delta$ for $i>1$ and $\beta_{i+1}-\beta_i = \Delta$ for $i<n$.
    
    The difference between the results from the above optimization schemes starts to be visible for $n \approx 8$. Hence, the two-parameter assumption is only a good approximation of the optimal pattern of displacements. Using it, we have improved our general optimization scheme: first, we perform a quick two-parameter optimization, repeating it a large number of times to avoid local minima. Then, we use the first stage result as a starting point for a single $2n-2$-parameter optimization, reaching the global optimum. 
    
\begin{figure}[ht]
		\includegraphics[width=\linewidth]{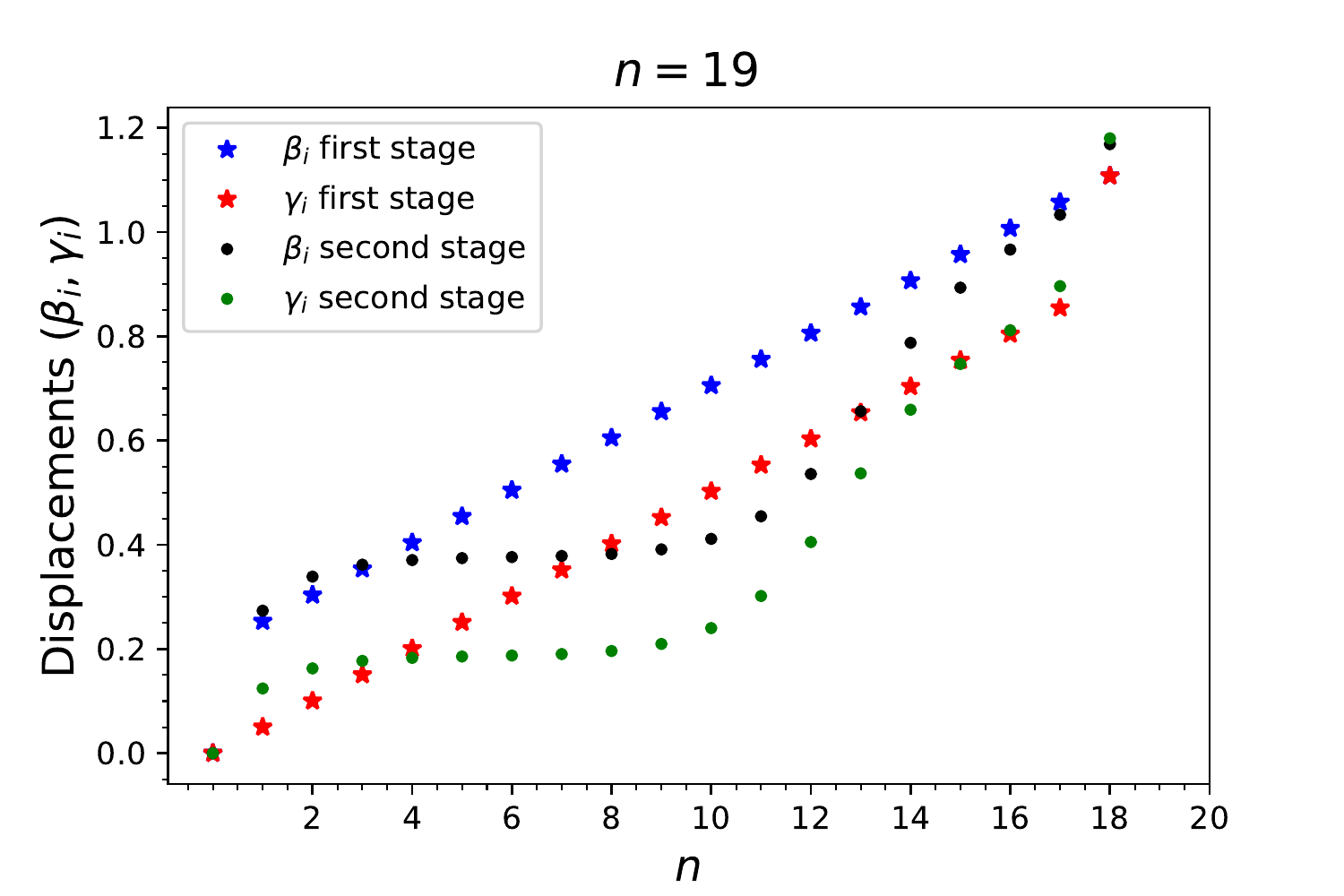}
		\caption{{\bf Comparison of the results of the first and second stage of optimization of the real displacements for $n=19$.} The Lab X displacements ($\{\beta_i\}$) and the Lab Y displacements ($\{\gamma_i\}$) after the first stage of optimization (marked by the blue and red start respectively) are arithmetic sequences except for values $\beta_1$ and $\gamma_n$. Black and green dots mark the values of $\{\beta_i\}$ and $\{\gamma_i\}$ respectively after the second stage of optimization.}
		\label{19_betas_gammas}
    \end{figure}

    After the first stage of the optimization the last $n-1$ displacements $\beta_i$ and the first $n-1$ displacements $\gamma_i$ form arithmetic sequences. In the second stage of the optimization, this linear dependence obtains a sine-like component. Fig.~\ref{19_betas_gammas} shows the comparison of the results of the optimization after the first and second stages for $n=19$. 
    
\begin{figure}[ht]
		\centering
		\includegraphics[width=0.8\linewidth]{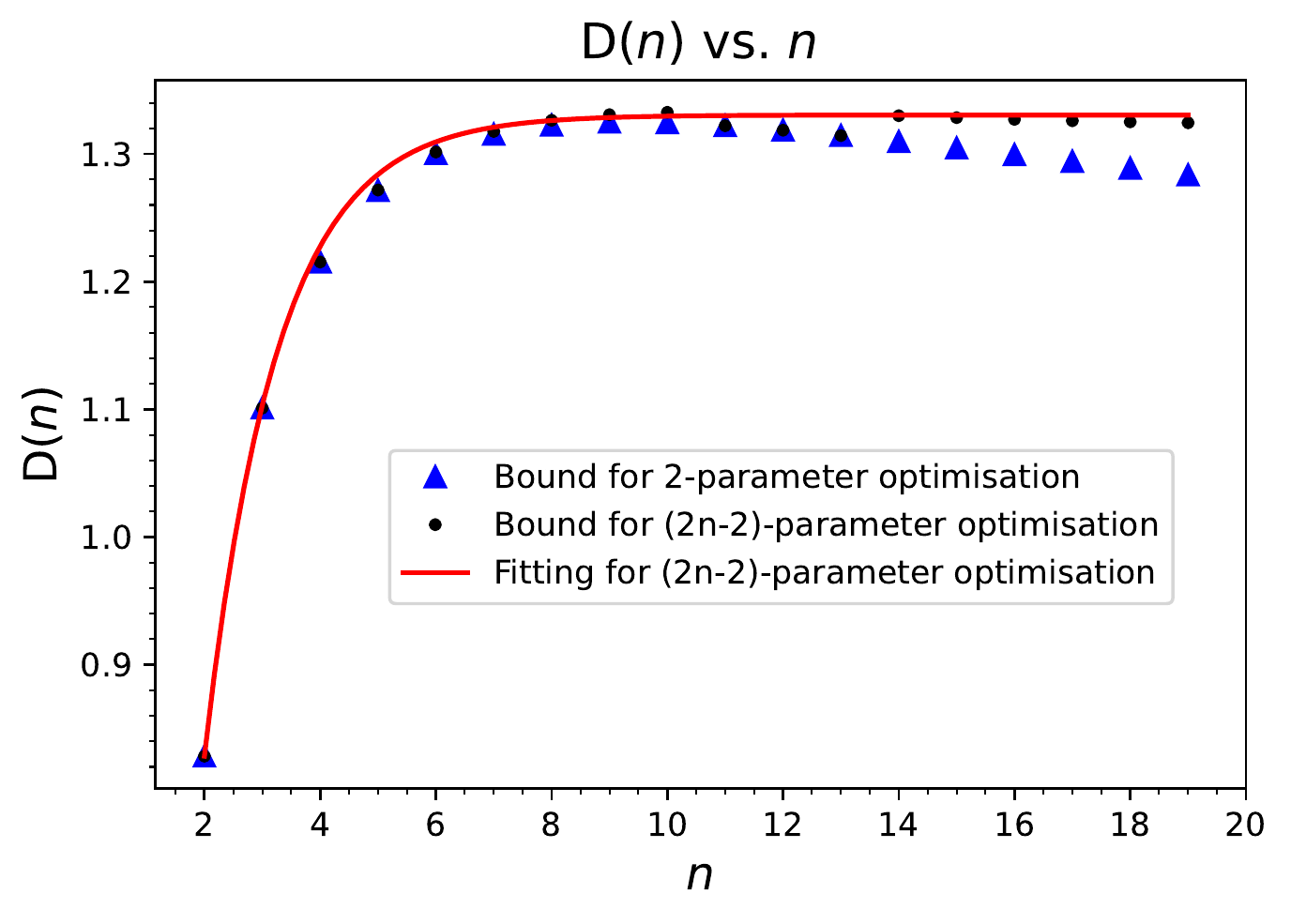}
		\caption{\textbf{Results of numerical optimization of maximal violation achievable by $n$ MZI settings:}
		The blue triangles represent the results of two-parameter optimization from multiple starting points. Starting from such obtained sets of displacements we proceed with a general, $2n-2$-parameter optimization and find the global optima, which values are represented by the black dots. 
		}
		\label{fig:two-stage}
    \end{figure}

    Fig.~\ref{fig:two-stage} shows the optimization results. The blue triangles are the first-stage (two-parameter optimization) results. The black dots are the results of the second stage. These values can be surprisingly well-fitted by an exponential function: $D(n) = C - A \exp(- B n)$. As the maximal violation for MZI settings is known for $n=2$ \cite{dastidar2022detecting} and equals $2\sqrt{2}-2$, we have two unknown parameters, $C$ and $B$: 
    \begin{equation}
        D(n) = 2\sqrt{2} - 2 + A ( \exp (-2B) - \exp(-B n) )
    \end{equation}
    In a non-linear regression, we obtain the values $A = -2.246$, $B = 0.7492$, and the correlation matrix entries: $S_A^2 = 0.0041$, $S_B^2 = 0.00024$, $Cov(A,B) = -0.0010$. The final form of the fitted exponent is:  
    \begin{equation}
        D(n) = 1.3377 - 2.246 \exp(-.7492 n) )
    \end{equation}
    
    \begin{figure}[ht]
		\centering
		\includegraphics[width=0.8\linewidth]{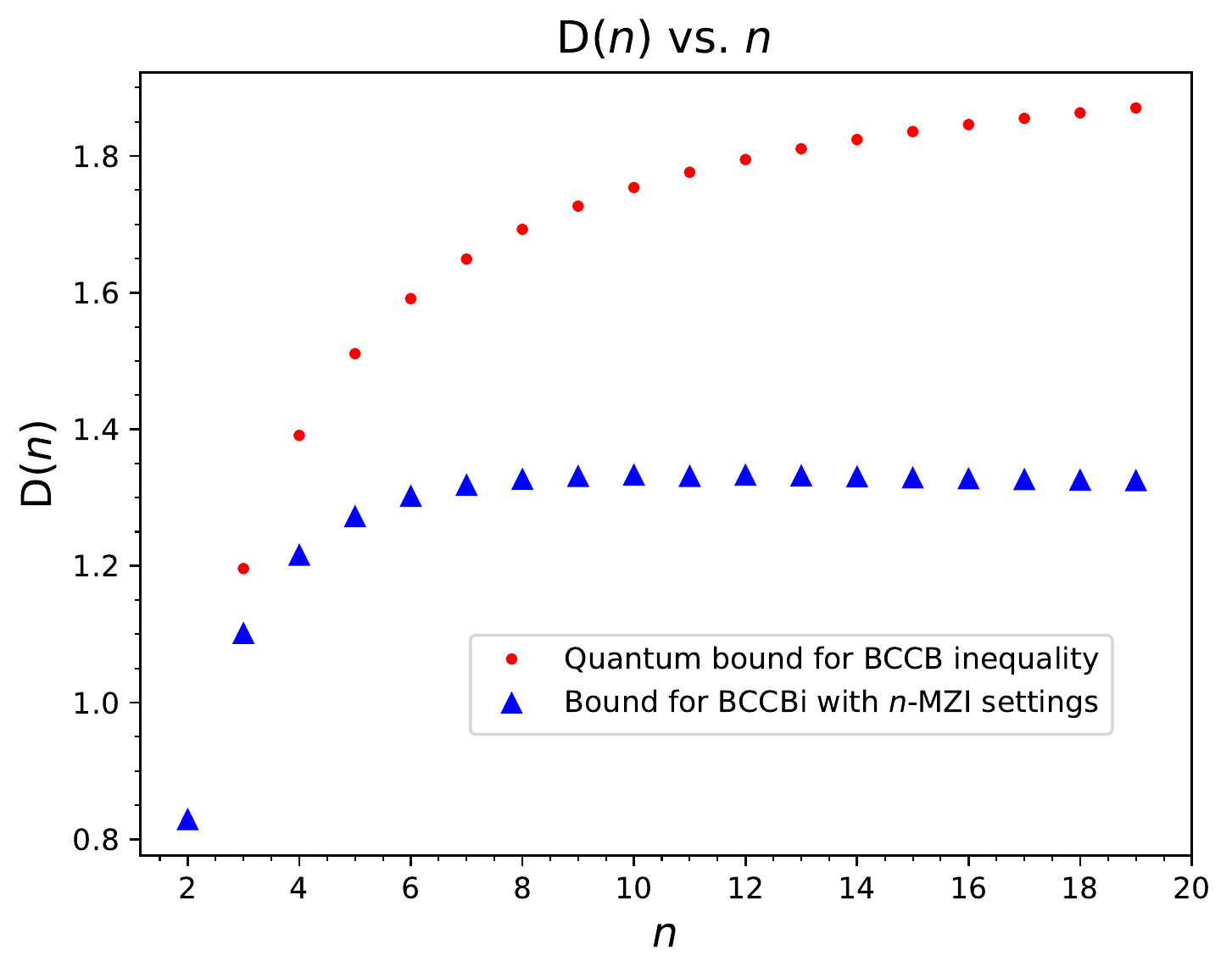}
		\caption{\textbf{Comparison of the general bound and that achieved by $n$ MZI settings:}
		Comparison of numerically generated plots for the maximum violation obtained by the BCCB inequality given in~\cite{wehner2006tsirelson} [in red] and that for $n$-MZI settings [in blue]. 
		}
		\label{max_viol_eigval}
    \end{figure}
    
    The Fig.~\ref{max_viol_eigval} compares the theoretical violation bound $\mathrm{D}(n) = \mathbb{E}(S)-2n+2$ (red points) with the numerical optimization results for $n \in [2, \dots, 19)$ (blue triangles).
    For $n>2$, we observe a significant difference between maximal violation ($D(n)$) and the violation achievable by $n$-MZI settings.
     
    For $n=2$, any two dichotomic observables $I-2\ket{\alpha_1}\bra{\alpha_1}, I-2\ket{\alpha_2}\bra{\alpha_2}$ generated by MZI settings can be represented as $I \otimes \sigma_1$ and $I \otimes \sigma_2$ on $\mathrm{span}\{\alpha_1,\alpha_2\} \oplus \mathrm{span}\{\alpha_1,\alpha_2\}^\perp$, where $\sigma_i$ have eigenvalues $\pm 1$ (and hence, are combinations of Pauli matrices). Choosing appropriate $\alpha_1 - \alpha_2$, one can obtain a commutation relation between $\sigma_1$ and $\sigma_2$ producing the maximal violation.
    
    On the other hand, for $n>2$, the observables generated by $n$ different MZI settings $\alpha_1, \dots, \alpha_n$ are necessarily linearly independent (due to linear independence of coherent state vectors), contrary to sets of Pauli matrices (\ref{XPauli}), (\ref{YPauli}), which span two-dimensional operator subspaces. This explains, why it is impossible to realize the violation-maximizing observable algebra for MZI settings. Also, it explains the gap between the maximal violation and the violation achievable by MZI settings for $n>2$.
    
    The maximal violation is related to sequences of displacements that are co-linear on the complex plane. On the contrary, let us assume, that we have no knowledge about the phase. The projector $\ket{\alpha}\bra{\alpha}$ has to be now averaged:
    \begin{align}
        \widetilde{P}_\alpha 
        & = \frac 1{2\pi} \int_0^{2\pi} \ket{e^{i\phi}\alpha}\bra{e^{i\phi}\alpha} \mathrm{d} \phi
        \nonumber \\
        & = \exp(-|\alpha|^2) \sum_{k=0}^\infty \frac {|\alpha|^2}{n!} \ket{k}\bra{k}.
    \end{align}
    The averaging decoheres the projector - kills off-diagonal (in the Fock basis) entries of its matrix. The projector becomes a positive operator and the projective measurement becomes a POVM, having both effects diagonal in the Fock basis. As all the $n$ POVMs commute now, the protocol is classical (all the measurements can see only the diagonal of the density matrix, not coherences) and thus, non-locality cannot be detected.
    
    We conclude, that a phase synchronization mechanism is necessary to detect a non-locality between modes in a two-mode state of light using Mach-Zender interferometers. Each local laser feeding the MZI with a strong coherent light has to be in phase with incoming mode, hence with the laser triggering the source. We have then both local lasers in phase with the triggering laser of the source, hence they have the same frequency $\nu$. To obtain the interference in MZIs, both modes of light must have the same frequency $\nu$.  

\begin{figure*}[ht]
		\centering
		\includegraphics[width=0.9\linewidth]{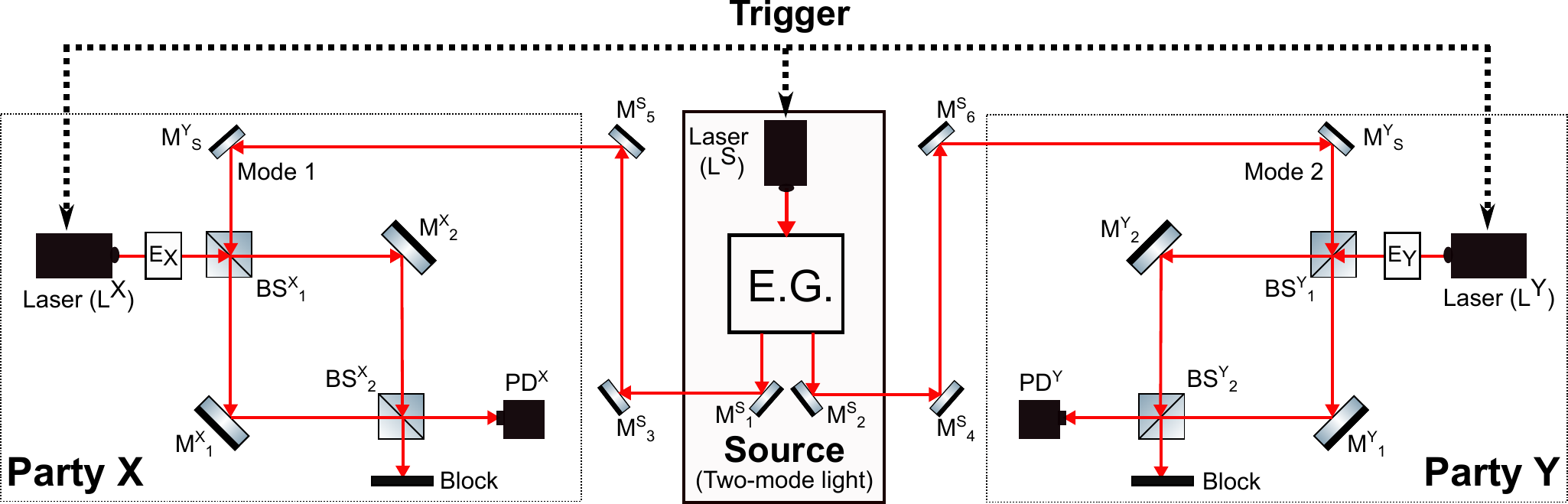}
		\caption{\textbf{Schematic of modified experimental setup: }The experimental setup is similar to that in Fig.~\ref{MZI_setup}. To make the phases equal for the input lights at both the MZI setups of party X and Y, we assume that two local lasers ($L_X$ and $L_Y$) are almost identical and the phase for both these laser lights can be tuned using electro-optic modulators $E_X$ and $E_Y$. Further, to match the frequencies of the source to that of the local laser, we assume that the source laser ($L_S$) is nearly identical to the two local lasers and is sent via an "entanglement unit" (E.U.) to generate the two-mode entangled light. Both the beams (if having down-converted frequencies such that they do not match that of $L_S$) can be up-converted to the original frequency of $L_S$ using U.C. (up-converting crystal). We do not discuss the generation two-mode entangled light as that is beyond the scope of our manuscript. However, the generation of ECS (a type of two-mode entangled light) can be found here~\cite{israel2019entangled}.}
		\label{expt_setup2}
    \end{figure*}

    In Fig.~\ref{expt_setup2}, we present 
    a scheme 
    to implement the phase synchronization mentioned earlier.
    For each lab, we assume that the local lasers are identical, thus having almost equal frequencies. 
    Further, tuning of the frequencies can be done by using an electro-optic modulator, to maintain 
    a constant relative phase between the lasers of both parties over time 
    ($E_Y$, see Fig.~\ref{expt_setup2}).
    Since in the practical scenario, each laser will have a frequency spread, the electro-optic modulators can be used to adjust and fix the phase of one laser w.r.t to the other. 
    Further, to have 
    both modes of the two-mode entangled state of light 
    being in phase with the respective local lasers, the laser pumping the source of the two-mode entangled state has to be synchronized with both local lasers. A similar synchronization scheme is described in~\cite{israel2019entangled}.
    We realize the source synchronization using another electro-optic modulator ($E_X$). An alternative for using the electro-optic modulators is to use one laser beam and divide it to feed MZIs in both laboratories and the source of the entangled state. This will not guarantee equal phases, but phase differences are constant in time, which is enough to satisfy the requirement of collinearity. 
    
    \subsection{States for Maximal Violation}\label{States for Maximal Violation}
    
    In the Sec.~\ref{Maximal Eigenvalues of CHSH matrix}, we have maximized the highest eigenvalue of $S$ (\ref{S_beta_gamma}) - the LHS of the BCCB inequality (\ref{genCHSH}) for $n = 2, \dots, 39$, obtaining its maximal expected value, resulting in maximal violation of the BCCB inequality.
    
    In this section, we discuss the structure of the pure, two-mode light states represented by the corresponding eigenvector of $S$.
    
    To verify our results, we check for the case $n=2$. The numerical model yields the eigenvector corresponding to the maximal violation, whose analytical expression is: 
    \begin{equation}
        \ket{\psi_2} = \frac{1}{\sqrt{2-\sqrt{2}}}\begin{bmatrix}
            -1\\
            1\\
            \sqrt{2}-1\\
            -1
        \end{bmatrix}.
    \end{equation}
    $\ket{\psi_2}$, unitarily equivalent to the maximally entangled state, is exactly what has been found analytically in~\cite{dastidar2022detecting}. Therefore, we proceed to comment on the states for $n>2$-settings.
    
    \begin{figure*}[ht]
		\includegraphics[width=\linewidth]{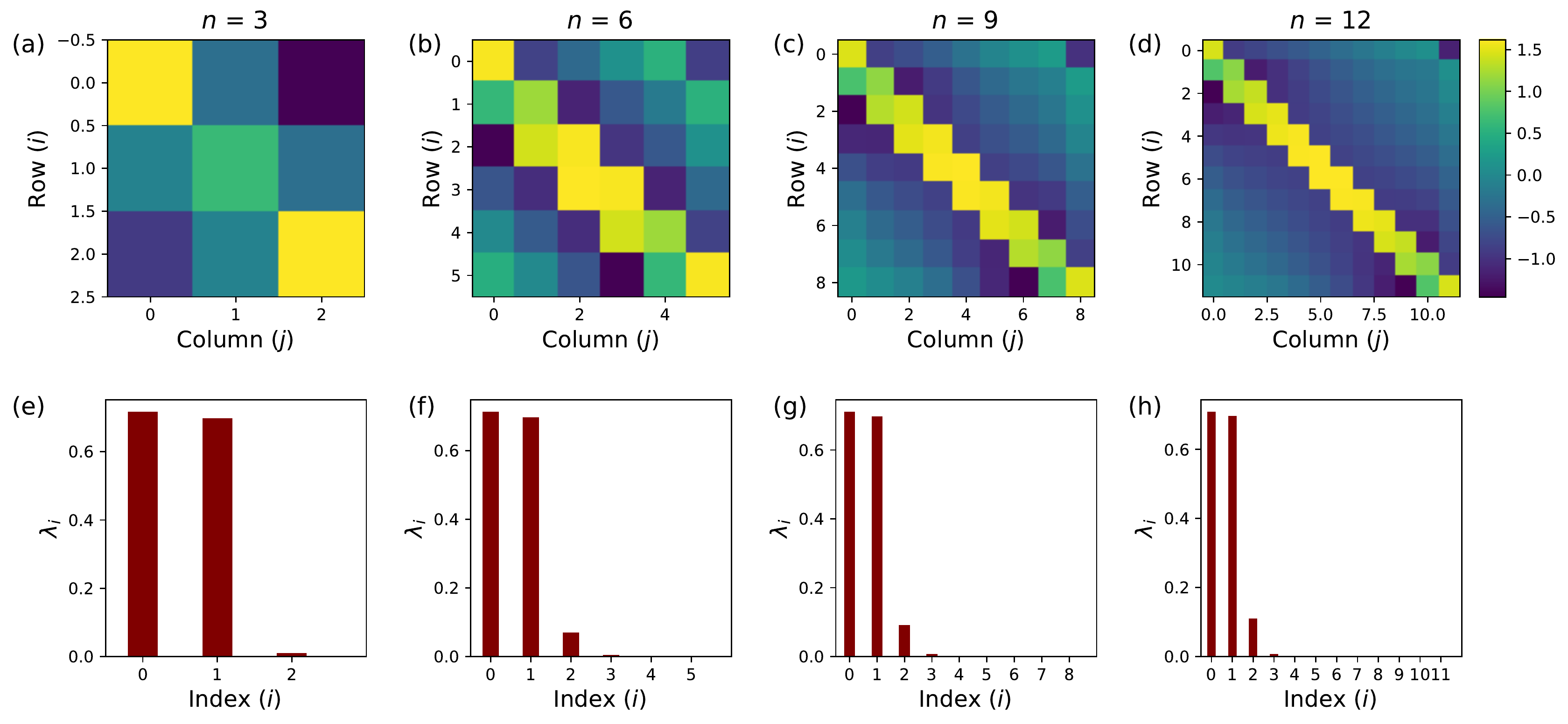}
		\caption{\textbf{Eigenvectors for maximum violation and their Schmidt coefficients:} The eigenvectors corresponding to the maximal violation. Vectors are represented in the (non-orthogonal) basis $\{\beta_i \otimes \gamma_j \}$ and their ($n^2\times 1$) arrays of coefficients are reshaped to $n\times n$ matrices and
		plotted for (a) $n=3$, (b) $n=6$, (c) $n=9$ and (d) $n=12$. The color bar on the right-end indicates the scalar values associated with the entries of the matrices plotted. (e)-(h) Bar graphs showing the magnitude of the Schmidt coefficients ($\lambda_j$, $j=1 \text{ to }n$) for the corresponding eigenvectors plotted above are shown.}
		
		\label{max_eigvec_schmidt_coeff_plots_real}
    \end{figure*}

    The entries of vectors $\ket{\psi_n}$ are real (see Remark \ref{S_real} in the Appendix \ref{A: Orthonormal Basis}). In the Fig.~\ref{max_eigvec_schmidt_coeff_plots_real}(a)-(d) we plot the values of entries of $\ket{\psi_n}$ in the non-orthogonal $\{\beta_i \otimes \gamma_j \}$ basis for $n = 3, 6, 9, 12$ after reshaping $(n^2\times1)$ column vectors to $(n\times n)$ matrices.
    
    Next we calculate the Schmidt coefficients~\cite{ekert1995entangled,nielsen2001quantum} of vectors $\Psi_n$ in the orthonormal (computational) basis, performing the singular value decomposition of the corresponding $(n \times n)$ matrices
    [see Figs.~\ref{max_eigvec_schmidt_coeff_plots_real}(e)-(h)].
    We observe that even for high $n$ the Schmidt rank of $\ket{\psi_n}$ is $4$ and the first two Schmidt coefficients dominate. 

    The BCCB inequality is maximally violated, if each party's observables are combinations of Pauli matrices (\ref{XPauli}, \ref{YPauli}), and then the maximal violation is realized for the singlet state. We have already commented, that for $n>2$ it is impossible to reconstruct such observables by displacement operators. Although, in the optimization procedure, the observable $S$ (LHS \ref{S_beta_gamma}) tries to resemble to the optimal one as much as possible, hence its eigenvector resembles the singlet state vector. 
    
    \section{Optimal Violation for Experimentally Achievable States}\label{Optimal Violation for Experimentally Achievable States}
    
    In the previous sections, we have discussed the maximum violation of the BCCB inequality achievable by $n$ MZI settings and calculated the state vectors for which such violation can be achieved.
    We have found that for $n=2$ the state vector for which the maximal violation is obtained can be written as~\cite{dastidar2022detecting}:
    
    \begin{align}\label{psi2}
    \ket{\Psi} = \frac{1}{\sqrt{2-\sqrt{2}}}\Big\{
    & \big[ \ket{\beta_1} - \ket{\beta_2} \big]
    \otimes
    \big[ \ket{\gamma_1} - \ket{\gamma_2} \big] \nonumber \\
    - & (2-\sqrt{2}) \ket{\beta_1} \otimes \ket{\gamma_1}
    \Big\},
    \end{align}
    where $\langle \beta_1 | \beta_2 \rangle = \langle \gamma_1 | \gamma_2 \rangle = 1/\sqrt{2}$.
    Similarly, for $n>2$, the maximal violation will be realized for a state vector from $\mathrm{span}\{\beta_1, \dots, \beta_n\} \otimes \mathrm{span}\{\gamma_1, \dots, \gamma_n\}$.
    
    However, experimental realization of such states and their applications have not been reported yet. Therefore, in this section, we will discuss how the BCCB inequality (\ref{genCHSH}) is useful 
    to detect entanglement 
    in certain classes of 
    experimentally viable states: entangled coherent (EC) state~\cite{israel2019entangled} and two-mode squeezed vacuum (TMSV) state~\cite{eberle2013stable}, 
    known for their importance for applications in quantum metrology. 
    We describe the numerics for this section in the Appendices~\ref{A:Entangled Coherent State} and~\ref{A:Two-Mode Squeezed Vacuum State}. Note that, in this section, we do not constrain the measurement settings $\{\beta_1,\dots,\beta_n\}$ and $\{\gamma_1,\dots,\gamma_n\}$ 
    to be real, but simply 
    perform a fresh optimization 
    of the maximal violation of BCCB inequality using $n$-MZI settings,
    without any initial assumptions.
    
    \subsection{Entangled Coherent States}\label{Entangled Coherent State}
    
    In this subsection, we maximize the violation of the BCCB inequality using $n$-MZI settings, 
    over a class of entangled coherent states, $\ket{\Psi_{EC}}$, which are of the form:
    
    \begin{equation}\label{ECS}
        \ket{\Psi_{EC}} = N_\alpha \left( a\ket{\alpha}\otimes \ket{0}+\ket{0}\otimes\ket{\alpha} \right)
    \end{equation}
    
    where $N_\alpha = 1/\sqrt{1+|a|^2 +2 e^{-|\alpha|^2} \Re(a))}$ is the normalization factor~\cite{israel2019entangled}. Note, that $\alpha$ can be made real by a local unitary transformation. 
    We maximize the expression $\bra{\Psi_{EC}(\alpha,a) } S(\vec \beta, \vec \gamma) \ket{\Psi_{EC}(\alpha,a)}$ w.r. to the real parameter $\alpha$ and complex parameters $a, \beta_1, \dots, \beta_n, \gamma_1, \dots, \gamma_n$. 
    
    \begin{remark}
        Observe that performing displacement operations in both subsystems $D(\eta) \otimes D(\epsilon)$, one can obtain a more general state        $N_\alpha(\ket{\alpha+\eta}\otimes\ket{\epsilon}+\ket{\eta}\otimes\ket{\alpha+\epsilon})$. 
        Local unitary operations 
        $\hat{D}(\epsilon)$ and $\hat{D}(\eta)$
        can be
        performed by using a Mach-Zehnder interferometer fed by a strong coherent state~\cite{dastidar2022detecting,windhager2011quantum}.
        We will not consider these more general states, as they are related to (\ref{ECS}) by a local unitary operation and have the same amount of entanglement.
    \end{remark}
    
    According to the Remark, as the canonical form (\ref{ECS}) of the state is fixed, we cannot reduce the number of parameters in the sequences of displacements by use of local unitary operations. Although, we observe no change in the optimal violations when we restrict ourselves to real displacements and real parameter $a$. Hence we can reduce the number of parameters in the optimization to $2n+2$.
    
\begin{figure}[ht]
		\centering
		\includegraphics[width=\linewidth]{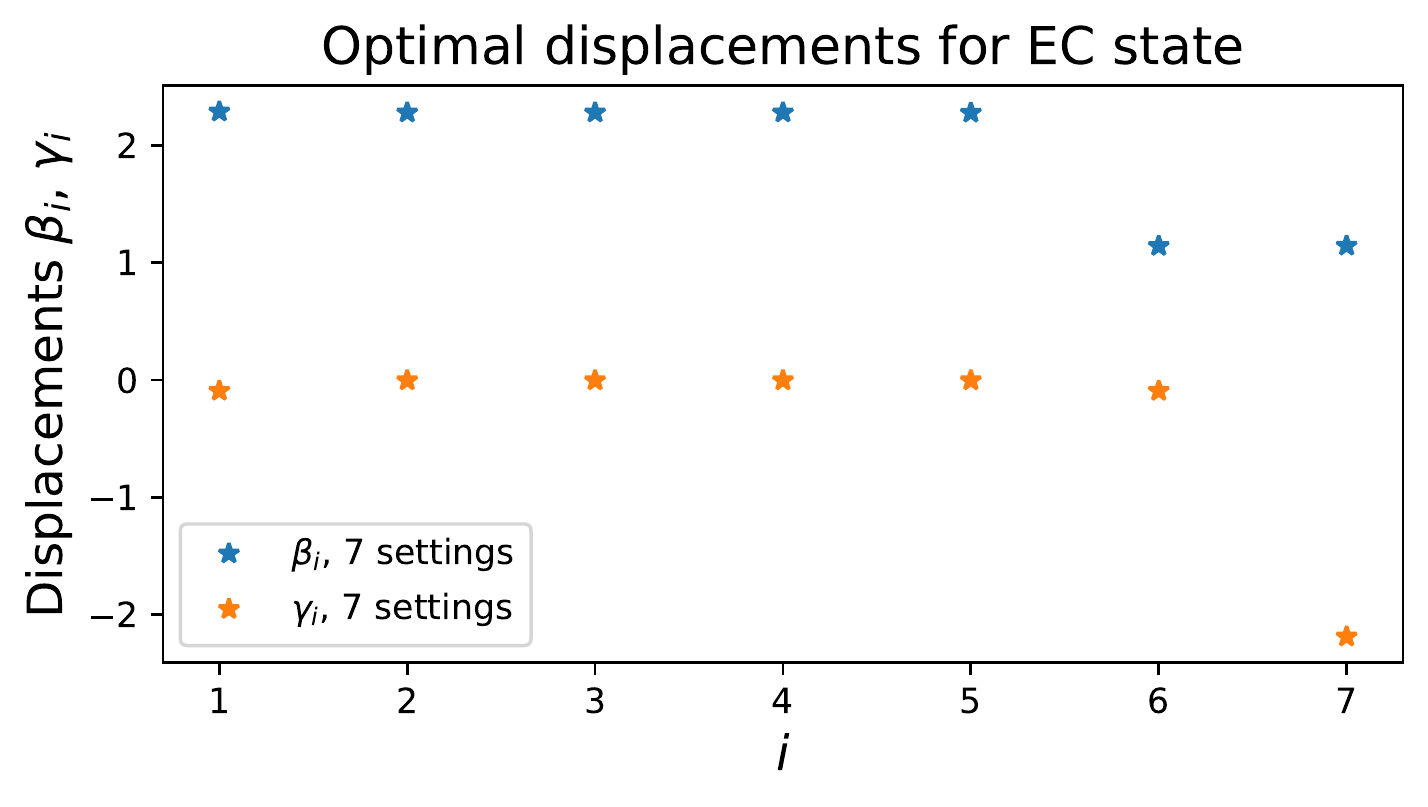}
		\caption{Real displacements optimizing the violation for EC states for $n=7$.
		}
		\label{ECS_displacements}
    \end{figure}

    A typical sequence of displacements for $n=7$ is shown in Fig.~\ref{ECS_displacements}.
    
    We observe that the displacements satisfy (approximately) the following relations:
    $\beta_1 = \dots = \beta_{n-2}, \beta_{n-1} = \beta_n, \gamma_{1} = \gamma_{n-1}, \gamma_2 = \dots = \gamma_{n-2}$.
    This observation let us reduce the number of optimization parameters to $7$. We repeat such reduced optimization $300$ times to avoid a stuck in a local optimum. We use the result as a starting point for the second step - the full parameter optimization. Proceeding in this way we reproduce the values of general optimization for small values of $n$ and improve the results for larger $n$, hence the simplified optimization produces effectively an approximation located in the attraction basin of the global optimum.
    
\begin{figure}[ht]
		\centering
		\includegraphics[width=\linewidth]{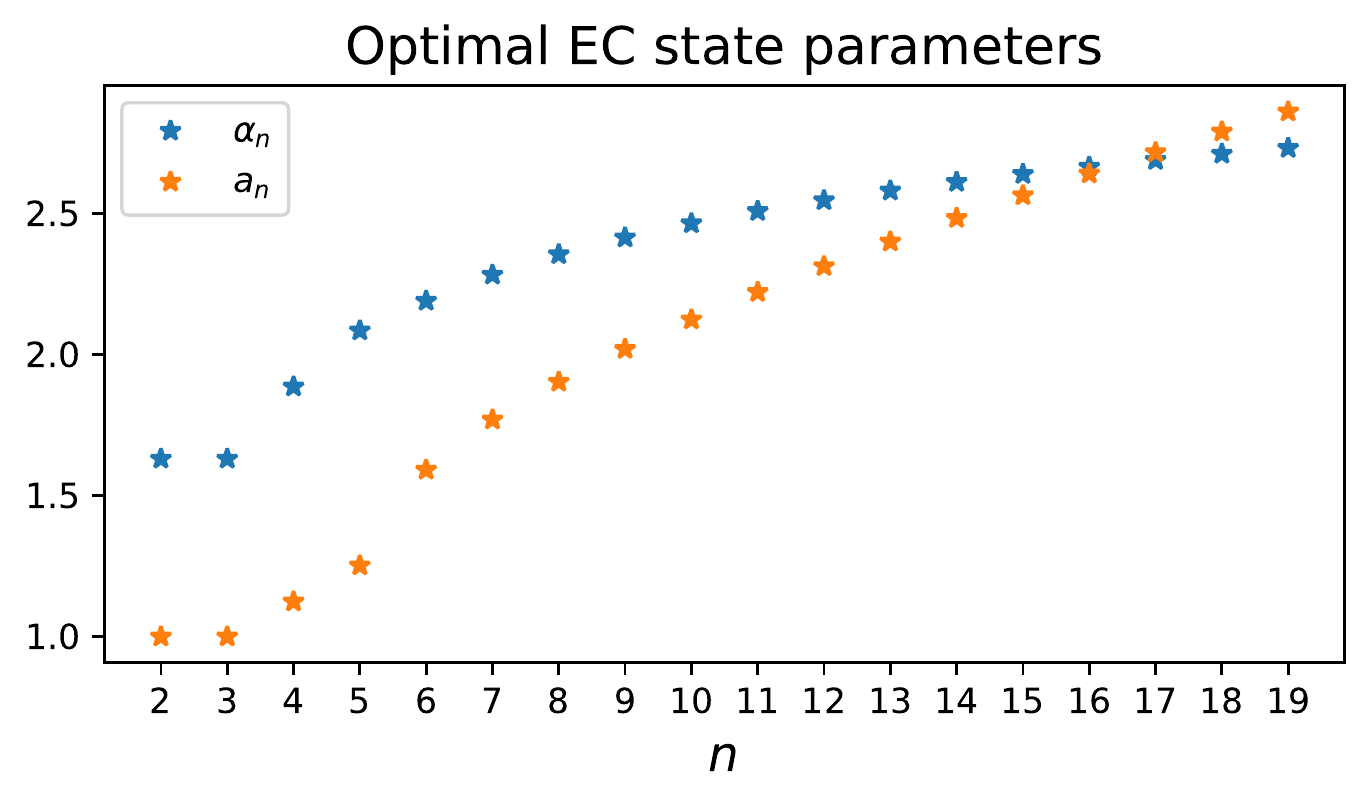}
		\caption{Parameters of entangled coherent state (\ref{ECS}) maximizing the violation of the BCCB inequality with MZI settings.
		}
		\label{ECS_parameters}
    \end{figure}

\begin{figure}[ht]
		\centering
		\includegraphics[width=\linewidth]{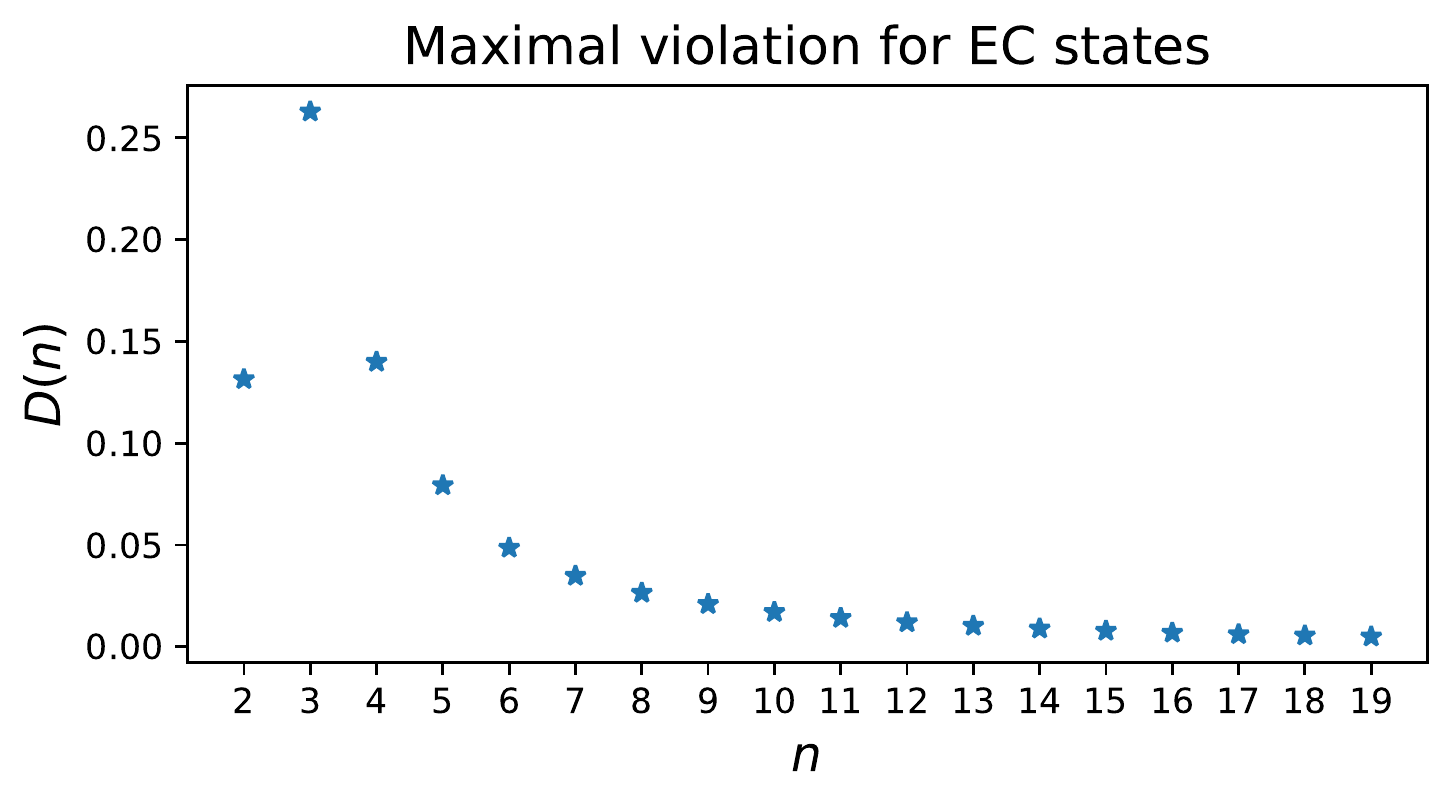}
		\caption{Maximal violation of the BCCB inequality with MZI settings in an entangled coherent state.
		}
		\label{ECS_viol}
    \end{figure}

    The parameters of EC states for which the maximal violation is realized are presented in Fig.~\ref{ECS_parameters}. Observe, that between values $6$ and $7$, we observe the change of the function behaviour - possibly two local optima exchange their role of the global optimum. In the numerical optimization, one can observe a frequent stuck in a local minimum for the value $6$. The maximal violation for EC states is shown in Fig.~\ref{ECS_viol}. The violation attains its maximal value $0.262887$ for $n=3$ and decreases to zero for higher $n$.

    The calculation details are described in the Appendix~\ref{A:Entangled Coherent State}.
    
    \subsection{Two-Mode Squeezed Vacuum States}\label{Two-Mode Squeezed Vacuum State}
   
    Another interesting class of experimentally accessible states 
    are 
    two-mode squeezed vacuum states~\cite{eberle2013stable}. Such states 
    can be achieved when the squeezing operator ($S(\xi)$) acts on two-mode vacuum ($\ket{0,0}$):
    
    \begin{equation}\label{TMSV}
       \ket{\Psi_{TMSV}} = S(\xi)\ket{0}\otimes\ket{0} = \exp{\xi^*\hat{a}\hat{b}-\xi\hat{a}^\dagger\hat{b}^\dagger}\ket{0}\otimes\ket{0},
    \end{equation}
    where 
    $\hat{a}, \hat{b}$ are photon annihilation operators in mode 1 and 2 respectively, and $\xi = re^{i\theta}$ is a squeezing parameter~\cite{gerry2005introductory}.
    The (\ref{TMSV}) can be expanded as:
    
    \begin{equation}\label{TMSV_decomposed}
        \ket{\Psi_{TMSV}(\xi)} = \frac{1}{\cosh{r}}\sum_{k=0}^\infty(-e^{i\theta}\tanh{r})^k\ket{k}\otimes\ket{k},
    \end{equation}
    where the phase $\theta$ is irrelevant, because it can be made $0$ by a local unitary $\exp(-i\hat{N}\theta/2)\otimes\exp(-i\hat{N}\theta/2)$, and hence we will assume $\xi = r \in \mathbb{R}_+$ since now.
    
\begin{figure}[ht]
		\centering
		\includegraphics[width=\linewidth]{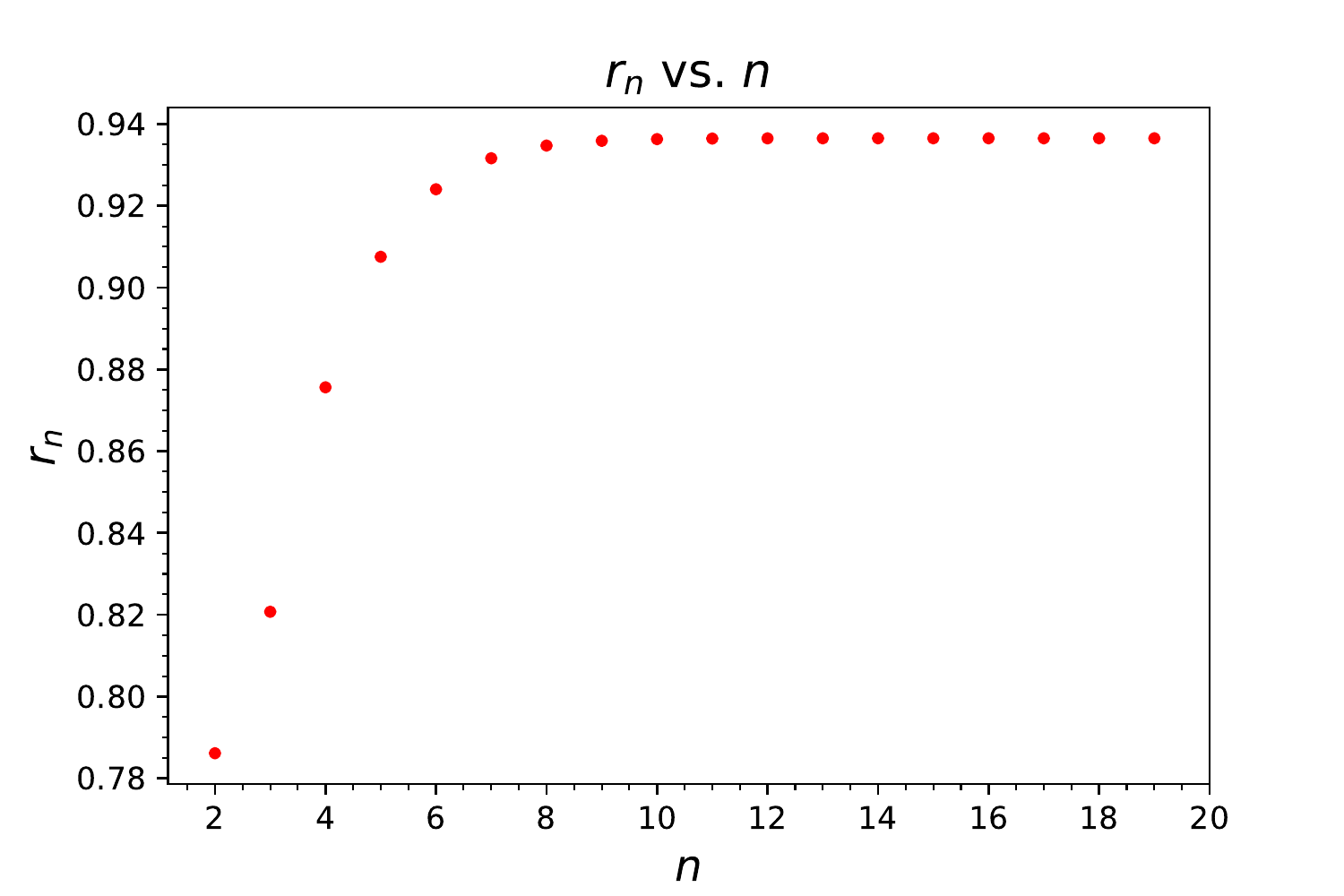}
		\caption{
		The value of squeezing parameter $r$, for which the maximal violation is obtained for $n$ MZI settings.
		}
		\label{r_vs_n}
    \end{figure}

\begin{figure}[ht]
		\centering
		\includegraphics[width=\linewidth]{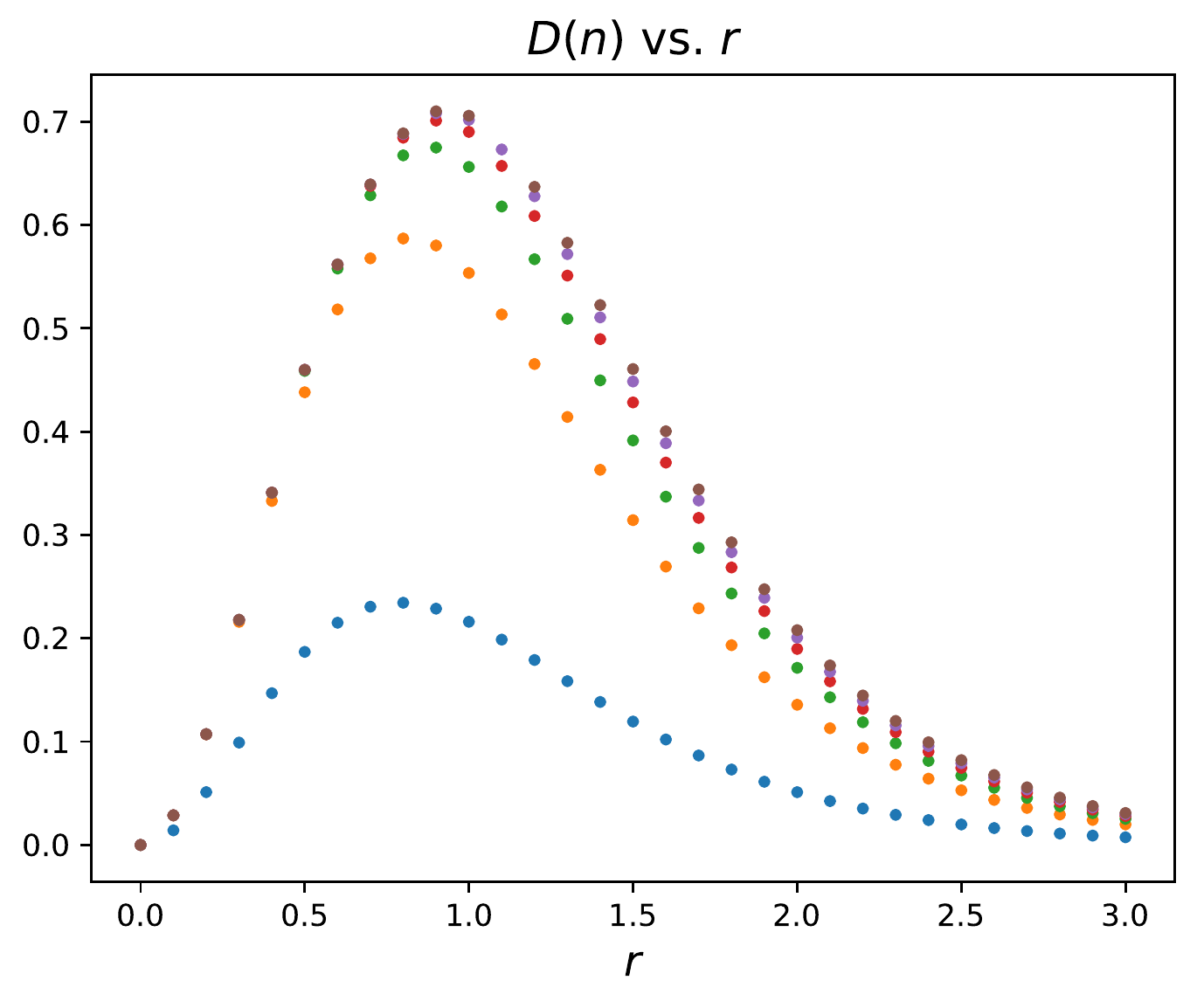}
		\caption{\textbf{Violation vs. squeezing parameter for different $n$ settings:} Plot for violation $D(n)$ vs. $r$ (squeezing parameter defining the TMSV state) varying from 0 to 3, for $n \in [2,7]$ being the number of MZI settings.
		}
		\label{tmsv_viol_r}
    \end{figure}

    Fig.~\ref{r_vs_n} shows the values of the squeezing parameter $r$ for which the maximal violation is obtained. Moreover, Fig.~\ref{tmsv_viol_r} shows the dependence of maximal violation on $r$ for different values of $n$ (for each value of $r$ an independent optimization is performed).
    
\begin{figure}[ht]
		\centering
		\includegraphics[width=\linewidth]{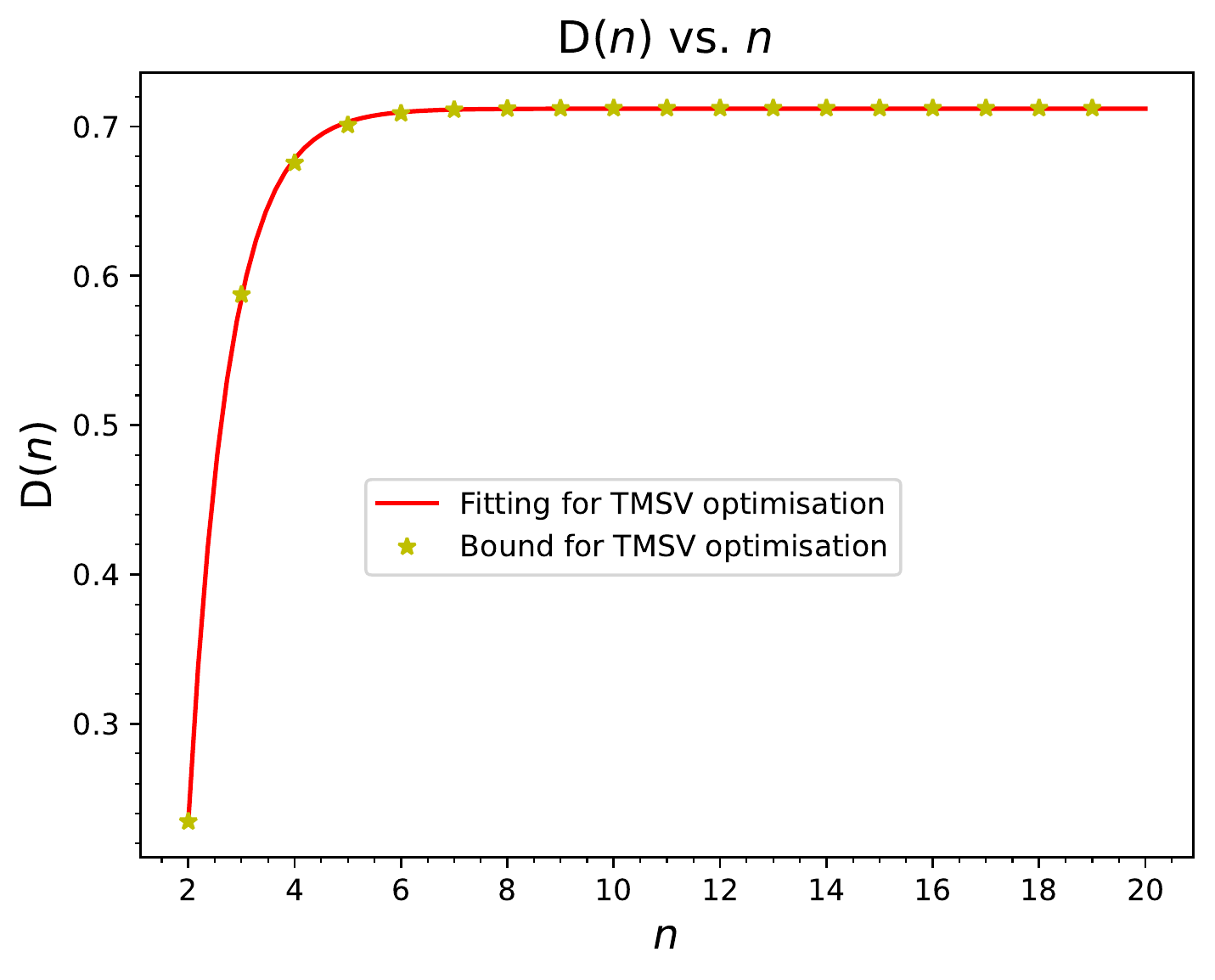}
		\caption{
		Values of maximal violation obtainable for $n$ MZI setups and a TMSV state (yellow stars). The red continuous line is an exponential saturation fitted to the data (more details in text).  
		}
		\label{tmsv_fitted}
    \end{figure}

    The dependence of maximal violation on $n$ is shown on Fig.~\ref{tmsv_fitted}. 
    We observe an exponential saturation of the value of maximal violation with the increasing $n$. We fit the function $a + c \exp(-bx)$ to the data. The fitted parameters are 
    $a = 0.71, b = 1.33, c = -6.79$
    and the corresponding covariance matrix is
    \begin{equation}
        C = \left[ \begin{array}{ccc} 
        7.97 \times 10^{-8} &  9.02 \times 10^{-6} & -7.41 \times 10^{-7} \\
       9.02 \times 10^{-6} &   1.28\times 10^{-2} & -9.06\times 10^{-4} \\
       -7.41 \times 10^{-7} & -9.06\times 10^{-4} &  6.52\times 10^{-5}
        \end{array} \right]
    \end{equation}

 
    
    
\begin{figure}[ht]
		\centering
		\includegraphics[width=\linewidth]{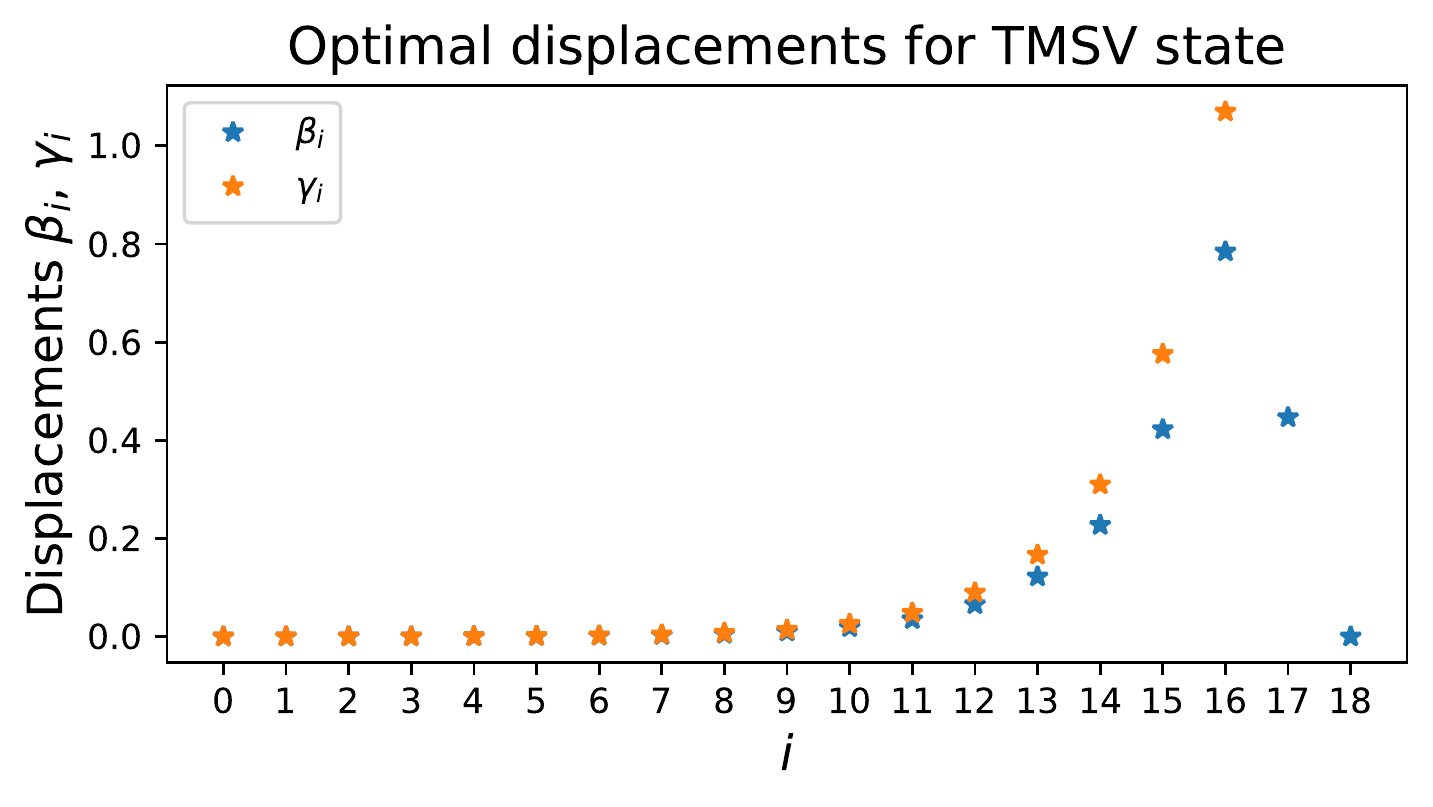}
		\caption{
		Values of displacements (real) in the MZI setups for $n=19$ optimizing violation for TMSV states. $\beta_i$ and $\gamma_i$ are displacements in MZI of Lab X and Lab Y respectively.  
		}
		\label{tmsv_dis}
    \end{figure}

    Restriction to real displacements in the optimization does not affect the values of maximal violation. The real displacements for $n=19$ shown in Fig.~\ref{tmsv_dis} present a typical pattern of displacements optimizing violation for TMSV states.
    
    \subsection{Comparison}\label{Observations}
    
\begin{figure}[ht]
		\centering
		\includegraphics[width=\linewidth]{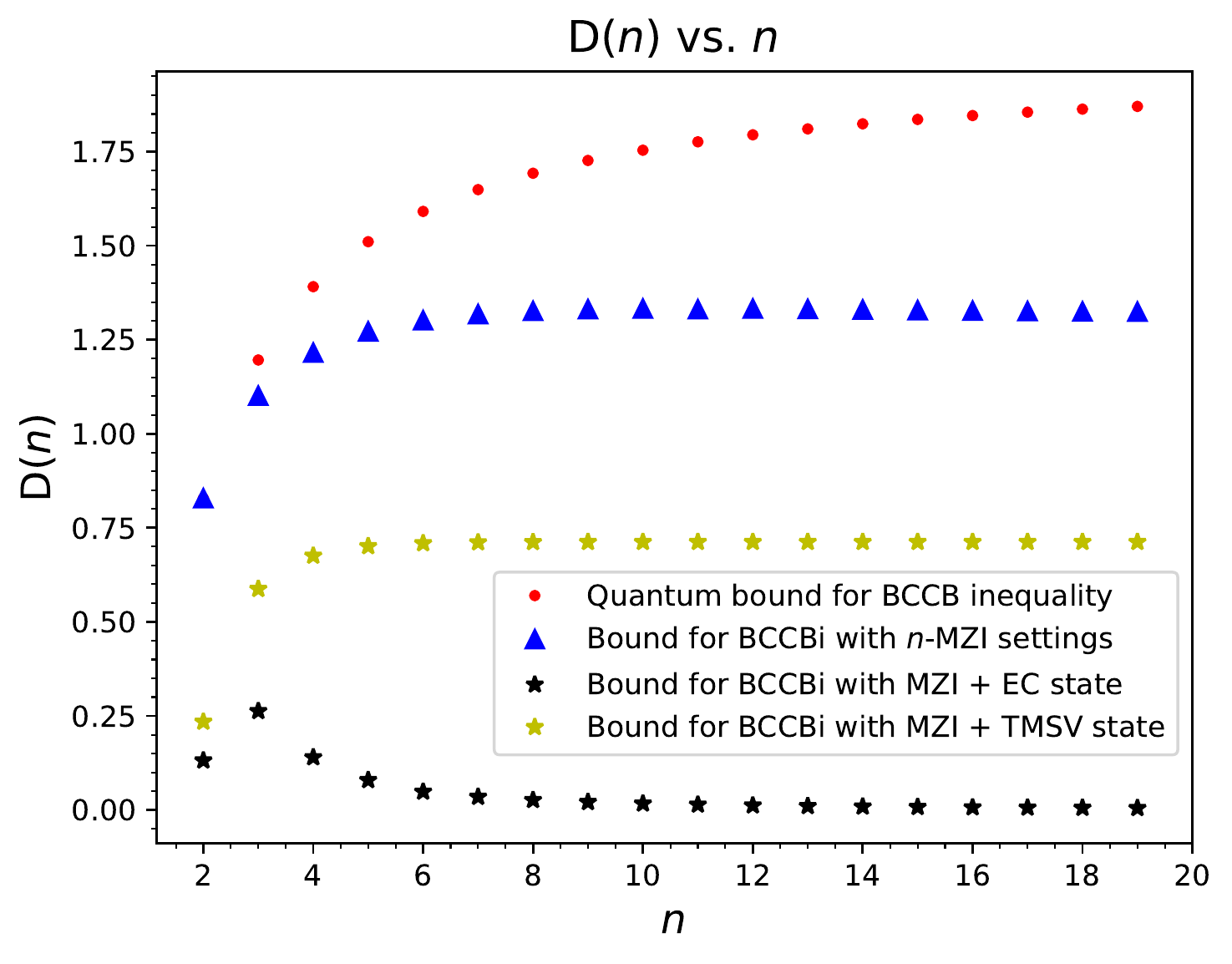}
		\caption{\textbf{Violations achieved by various protocols:} Comparison of numerically generated plots for the maximum violations obtained by the BCCB inequality proposed in~\cite{wehner2006tsirelson}, BCCB inequality described for $n$-MZI settings [in blue], MZI+ECS setting [in black] and MZI+TMSV [in yellow]. The plots have been generated for $n \in [2, 20)$.
		}
		\label{max_viol_all}
    \end{figure}

    In the following section, we compare and discuss the results of optimization for EC and TMSV states families and compare them with the results of optimization without state restriction and with the general violation bound for BCCB inequality. 
    The comparison is presented in Fig.~\ref{max_viol_all}.
    
    We observe that the violation of the generalised inequality (\ref{genCHSH}) can be observed for both EC and TMSV as input states when each of the two labs uses an MZI+photodetector arrangement, and that the maximally obtainable violations are much better for TMSV states than for EC states.
    The violation quickly becomes negligible (for $n > \approx 10$) for ECS. For TMSV, higher violations are obtainable and the maximal violation saturates to $\approx 0.7$.
    It is still significantly less than the violation achievable by the MZI setup when no restriction on states is given, which in turn is less than a theoretical bound, achievable when no restrictions of observables are assumed, provided by~\cite{wehner2006tsirelson}. 
    
     \section{Conclusion}\label{Conclusion}
    
    In summary, we have considered 
    the BCCB inequality 
    for detecting entanglement between two modes of light, when both parties use 
    $n$-Mach-Zehnder interferometric measurement settings 
    realizing 
    dichotomic observables. 
    We have 
    numerically optimized 
    the maximal violation obtained for this system and compared it with the theoretical bound, as in~\cite{wehner2006tsirelson}. 
    We observed that the violation achievable by $n$-MZI settings do not saturate the theoretical bound given by~\cite{wehner2006tsirelson}, for $n>2$. We provided justification for this fact: The bound is saturated for the singlet state of the qubit and for the involved observables being combinations of $\sigma_x$ and $\sigma_z$. On the other hand, different settings of MZI result in linearly independent observables, hence using MZI we can only approximate the optimal algebra of observables.
    
    The violation obtainable by MZI setups on both parties quickly saturates to the constant value of 1.3377.
    
    Next, we have restricted ourselves to two significant, experimentally accessible families of states:
    entangled coherent states and two-mode squeezed vacuum states. 
    We were able to violate the BCCB 
    inequality for both families using MZI+photodetector setups in each lab.
    For EC states, the maximal violation can be achieved for TMSV states, we see that the violation grows and reaches a constant value of approx. $0.7$ for $n\ge 4$. Thus, our experimental settings detect the entanglement in TMSV states better than that in ECS.
    
    In each optimization task, the maximal violations are obtained for real displacements. To guarantee a constant phase between a mode entering the MZI and the coherent laser field in the second input, one has to equip the experimental setup with a phase synchronization mechanism. We have proven, that no violation can be obtained if phases are not synchronized.  

    \acknowledgements

GS was supported by the National Science Centre project 2018/30/A/ST2/00837.
MGD was supported by the Prime Minister's Research Fellowship (PMRF), India. VPB acknowledges the financial support from DST QUEST grant DST/ICPS/QuST/Theme-2/Q35 and the Institute of Eminence scheme at IIT Madras, through QuCenDiEM group. 
    
%
    
    \onecolumngrid
    
    \vspace{.5cm}
    
    \appendix
    
    \twocolumngrid
    
    In the following Appendices, we will refer to codes written in Python programming language and using numerical packages \texttt{numpy}, \texttt{scipy} and \texttt{matplotlib}. The codes are
    accessible in a GitHub repository \url{https://github.com/madhuragd/generalization-of-CHSH}. 
    
	\section{Finite representation of algebra of observables}\label{A: Orthonormal Basis}
	
	One-party observables $A(\beta_i) = I - 2\ket{\beta_i}\bra{\beta_i}$ act non-trivially on the finite-dimensional subspace $\mathrm{span}\{\beta_1,\dots,\beta_n\}$ of the one-party Hilbert space. Although the pairwise different coherent state vectors $\{\beta_1,\dots,\beta_n\}$ are linearly independent and establish a proper basis of the finite representation, we need to write down the observables $A(\beta_i)$ in an orthonormal basis, because the spectrum of a matrix is invariant on unitary transformations. In this appendix, we focus on obtaining appropriate bases to represent $A(\beta_i)$ and $A(\gamma_i)$.
	
	An obvious way would be to obtain an orthonormal basis of $\mathrm{span}\{\ket{\beta_i}\}$ via the Gram-Schmidt ortho-normalization~\cite{leon2013gram}. For the $n=2$ case, we easily perform it to get:
	
	\begin{align}\label{OrthoBasis_n2}
	\ket{e_1} &= \ket{\beta_1}, \nonumber\\
	\ket{e_2} &= \frac{\ket{\beta_2} - \ket{\beta_1}\braket{\beta_1}{\beta_2}}{\sqrt{1 - \exp(-|\beta_1-\beta_2|^2)}}.
	\end{align}
	(see~\cite{dastidar2022detecting} for details). However, it is easy to check that for $n\ge2$, this direct method becomes cumbersome. Thus, we now proceed with the following algorithm to obtain an orthonormal basis from $\{\ket{\beta_1}, ..., \ket{\beta_n}\}$.
	
	We first arrange $\{\ket{\beta_i}\}$ into a $\infty \times n$ matrix $B=[\ket{\beta_1}| \ket{\beta_2}| ...| \ket{\beta_n}]$ such that the $i$th column of $B$ is $\ket{\beta_i}$ written in the standard (Fock) basis. Next, we construct an $n \times n$ Gram matrix ($G$) of $B$, i.e., $G = B^\dagger B$.
	We know that the elements of $G$ are 
	\begin{equation}\label{Gij}
        G_{ij} = \braket{\beta_i}{\beta_j} = e^{-(\abs{\beta_i-\beta_j}^2 + \beta_i\beta_j^* - \beta_j\beta_i^*)/2}.
	\end{equation} 
	
	Now, the Cholesky decomposition of $G = L L^\dagger$, where $L$ is a lower triangular $n \times n$ matrix with real and positive diagonal entries. It can be easily seen that $(BL^{-1\dagger})^\dagger(BL^{-1\dagger}) = \mathbb{I}_n$. Thus, the columns of $\infty \times n$ matrix $BL^{-1\dagger}$ are an orthonormal basis of $\mathrm{span}\{\ket{\beta_1}, \dots, \ket{\beta_n} \}$. 
	Let us call the vectors corresponding to these columns by $\ket{e_1}, ..., \ket{e_n}$. The one has:
	
	\begin{eqnarray}
	 &BL^{-1\dagger} = [\ket{e_1}| \ket{e_2}| ...| \ket{e_n}] \text{ and,} \nonumber \\
	 &B = [\ket{\beta_1}| \ket{\beta_2}| ... |\ket{\beta_n}] = [\ket{e_1}| \ket{e_2}| ...| \ket{e_n}]L^{\dagger}
	\end{eqnarray}
	
	Thus the $i$th column-elements of $L^\dagger$ are the coefficients of $\ket{\beta_i}$ in the orthonormal basis $\{\ket{e_i}\}$.
	Let $\{ \ket{f_i} \}_{i=1}^n$ denote the standard basis of $\mathbb{C}^n$. One picks the $i$-th column of a matrix multiplying if by $\ket{f_i}$ from the right. Hence  $\ket{\beta_i} = L^\dagger\ket{e_i}$ and we have:
	\begin{equation}\label{Abeta_ei}
	    A(\beta_i) = \mathbb{I} - 2L^\dagger\ketbra{f_i}L
	\end{equation}
	
	Similarly, for Lab Y, we can repeat the same procedure to find the orthonormal basis from $\{\ket{\gamma_i}\}$. Doing so, we obtain $\ketbra{\gamma_i} =  K^\dagger\ketbra{f_i}K$ where $H = K^\dagger K$ is the Cholesky-decomposed Gram matrix ($H$) corresponding to $\ket{\gamma_i}$. 
	
	The procedure is implemented in the function \texttt{genL} in the file \texttt{util.py}.
	For an array of displacements, given as the argument, the function first calculates the Gram matrix \texttt{G} of the corresponding coherent state vectors with entries given by (\ref{Gij}) and next it calculates the matrix \texttt{L} of the Cholesky decomposition of \texttt{G} using \texttt{numpy.linalg.cholesky}.
	
	During the procedure, an exception
	\texttt{np.linalg.LinAlgError(Matrix is not positive definite)} sometimes occurs.
	This happens because during the calculations some diagonal elements become negative due to numerical inaccuracy. If during the minimization procedure, two displacements become close to each other such a situation may arise.
	For handling this exception, we add a correction factor (3 $\times$ modulus of the smallest eigenvalue of $G$) to the diagonal terms in the matrix. This trick vastly reduces the exception occurrence frequency, but does not guarantee success - exception occurs also if positive eigenvalues are too small, hence the whole function should be used in an exception-handling block.
	
	Next the function \texttt{one\_party\_local\_observables} provided with an array of displacement constructs finite-size matrices representing observables $A(\beta_i)$. It uses the previously described function \texttt{genL} to generate the matrix \texttt{L} and then construct the matrices of observables from projectors onto conjugated rows of the matrix $L$ (columns of $L^\dagger$).
	
	\begin{remark} \label{beta_positivity}
	    In case all $\{\beta_i\}$ are real, $L=L^\dagger$.
	\end{remark}
	\begin{remark} \label{S_real}
	    In case all $\{\beta_i\}$ and $\{\gamma_i\}$ are real, the matrix $S$ given by (\ref{S_beta_gamma}) is a real symmetric matrix, and its eigenvectors have real 
	    entries.
	\end{remark}
	
    \section{Maximal Violation of BCCB inequality}
    
    \subsection{General states}\label{A: n-MZI}
    
    Once we have defined the function \texttt{one\_party\_local\_observables}, prescribing $n\times n$ matrices to the observables $A(\beta_i)$,
    for a given set of displacements $\{\ket{\beta_1}, \dots, \ket{\beta_n} \}$, 
    we define a function \texttt{LHS\_of\_BCCB} (in the file \texttt{util.py}) returning the $n^2 \times n^2$ matrix S (\ref{S_beta_gamma}) 
    for arrays of complex displacements $\{\beta_i\}$ and $\{\gamma_i\}$ given as its arguments.
	
	The function \texttt{max\_viol} in the file \texttt{Eig\_general.py} translates a real array of size $4n$ to $n$-dimensional vectors $\beta$, $\gamma$ of complex displacements, calculates the corresponding matrix $S$ using the \texttt{LHS\_of\_BCCB} function and returns the negative of its maximal eigenvalue. The function is passed as an argument to the function \texttt{scipy.optimize.minimize}, which finds its minimum using the Powell algorithm. 
	We start from random sequences of displacements and repeat the procedure a number of times to minimize the probability of getting stuck in a local minimum. 
    In this way, we obtain the maximal violation for the BCCB inequality using $n$-MZI settings.
    
    Now, for $n \in \{3, \dots, 8\}$, we obtain the sequences $\{\beta_i\}$ and $\{\gamma_i\}$ for which the violation is obtained under such a protocol, and plot them in the complex plane (Fig.~\ref{complex_plane_beta_gamma}). The whole code is under \texttt{Eig\_general.py}. The pickled results of optimization are in the file \texttt{Eig\_general.pi}.
    
    Next, we observe the collinearity of the resulting displacements on the complex plane. Hence by a local displacement operator and phase rotation, the sequences can be made real and the first displacement can be fixed to zero. Performing optimization over the reduced number of parameters we observe no decrease of maximal violation, hence the optimization can be performed over $2n-2$ real parameters and remains stable for higher $n$. The optimization is realized by the code \texttt{Eig\_real.py} and the results are stored in \texttt{Eig\_real.pi}. The only modification in comparison to the previous code is how the function \texttt{max\_viol} translates the array of $2n-2$ real numbers to sequences of displacements $\{\beta_i\}$, $\{\gamma_i\}$.
    
    Differences between subsequent displacement in the sequences shown in Fig.~\ref{complex_plane_beta_gamma} are almost equal, except the first/last one being significantly bigger. As we have already discussed, such simplified $4$ parameter optimization is only approximate but leads to a point in the attraction basin of the global (calculated by $2n-2$ parameter optimization) optimum. We use this observation to perform many times the fast simplified optimization to avoid stuck in a local minimum and then only once the full $2n-2$ parameter optimization. The first step is implemented in the file \texttt{Eig\_first\_stage.py} and the results are serialised in \texttt{Eig\_first\_stage.pi}. The serialised data is loaded in the code \texttt{Eig\_second\_stage.py} and the results are stored in \texttt{Eig\_second\_stage.pi}.
    
    The code \texttt{Eig\_graphs.py} produces a figure of plots of displacements for $n = 3 \dots 8$ using the data from the file \texttt{Eig\_general.pi} and the plot of maximal violations w.r.t. $n$ using the data from the file \texttt{Eig\_second\_stage.pi}.
    It fits the exponential decay to the data, plots the fitting and prints the values of parameters and the covariance matrix. 
    
    Next, we analyse the pure states realizing maximal violation using 
    the the results pickled in \texttt{Eig\_second\_stage.pi}.
    To do this, 
    the code \texttt{Eigvectors\_max\_violation.py} recovers the eigenvector corresponding to the maximal eigenvalue of the matrix $S$ (\ref{S_beta_gamma}) for each $n$ using \texttt{numpy.linalg.eig}.
    For each eigenvector, we calculate its decomposition in the (non-orthogonal) basis $\{ \beta_i \otimes \gamma_i \}$ of coherent-state vectors. A dictionary, prescribing to each $n$ both decompositions is stored in \texttt{max\_viol\_states.pi}.
    
    The code \texttt{Plots\_Eigvecs\_SchmidtCoeffs.py} loads for each $n \in \{ 3, 6, 9, 12\}$ two decompostions of the correcponding eigenvector from \texttt{max\_viol\_states.pi} and reshapes them to $n\times n$ matrices. The entries of the coherent state vector decomposition are plotted in the top row of the figure. The bottom row of the figure presents the plots of Schmidt coefficients of eigenvectors calculated using Singular Value Decomposition (\texttt{numpy.linalg.svd}) on the $n\times n$ matrix of coefficients in the orthonormal basis.
    
    \subsection{Entangled Coherent States}\label{A:Entangled Coherent State}
    
    The expectation value of the observable $S$ (\ref{S_beta_gamma}) w.r.t. the particular state vector $\ket{\Psi_{ECS}}$ is:
    \begin{widetext}
    \begin{align}\label{E_ECS}
        \mathbb{E}(S)_{ECS} & = \bra{\Psi_{ECS}} S\ket{\Psi_{ECS}} = \bra{\Psi_{ECS}}\Big(\sum_{i=1}^n A(\beta_i) \otimes A(\gamma_i) + \sum_{i=1}^{n-1} A(\beta_{i+1}) \otimes A(\gamma_i) - A(\beta_1) \otimes A(\gamma_n)\Big)\ket{\Psi_{ECS}} \nonumber \\
        & = \sum_{i=1}^n f_{i,i} + \sum_{i=1}^{n-1} f_{i+1,i} - f_{i,n},
    \end{align}
    where 
    \begin{align}\label{f_ij}
        f_{i,j} = \bra{\Psi_{ECS}} A(\beta_i) \otimes A(\gamma_j)\ket{\Psi_{ECS}} = N_\alpha^2 \big[|a_1|^2b(\alpha,\alpha,\beta_i)b(0,0,\gamma_j)+2Re\big(a_1^*b(\alpha,0,\beta_i)b(0,\alpha,\gamma_j)+b(0,0,\beta_i)b(\alpha,\alpha,\gamma_j)\big)\big],
    \end{align}
    and $b(x,y,z) = \braket{x}{y} - 2\braket{x}{z}\braket{z}{y}$, $\braket{x}{y} = e^{-\abs{x-y}^2/2+\im Im(x^*y)}$,
    \end{widetext}
    where we have used the fact than $A(\beta) = \hat{D}(\beta)A(0)\hat{D}^\dagger(\beta)$ and the properties of displacement operators~\cite{gerry2005introductory}.
    
    We perform a numerical optimization of $\mathbb{E}(S)_{ECS}$ for each $n = 2, \dots, 19$ in the code \texttt{ECS.py} to find the maximal violation in this class of states. 
    
    The factory function \texttt{bccb}, prescribes a function calculating $\mathbb{E}(S)_{ECS}$ of an array \texttt{x}, for two arguments: a parametrizing function \text{param} and \texttt{n} - the number of MZI settings. The parametrizing function defines how to decode from an array \texttt{x} subsequent objects: \texttt{a} - superposition parameter, \texttt{alpha,b,c,d}-displacements in the state and \texttt{beta, gamma} - sequences of displacements realized by MZIs. 
    We use local unitaries to fix \texttt{b,c,d} to zero.
    
    Next we define the subsequent functions: \texttt{param1}, \texttt{param2}, \texttt{param3}, \texttt{param4} defines different parametrization functions, ordered due to increasing number of parameters. Each such function has a prescribed attribute \texttt{size} storing the dimension of the parameter space.
    
    In the main function \texttt{max\_violation} we minimize a function produced by the factory function \texttt{bccb}. The starting point is a random array of the size defined in the \texttt{size} attribute of the chosen function \texttt{param}. We repeat the minimization \texttt{m} times, to avoid stuck in a local minimum and then we use the result as a starting point of the full-parameter parametrization (we observed first, that it is enough to consider only real values of parameters). 
    
    It is enough to perform on the first stage a $5$-parameter optimization, defined in the function \texttt{param1}. It is not well-defined For $n=2$ and then we choose the full-parametrization \texttt{param3} (in this dimension they coincide). For $n=6$ the procedure does not lead to the optimal value (the result of the simplified optimization is not in the attraction basin of the global optimum) and we need to perform a more detailed parametrization \texttt{param2}.   
    
    The results are pickled to the file \texttt{max\_viol\_ecs.pi}. The data is then used by the code \texttt{ECS\_graphs.py} to produce graphs of optimal values of state parameters, optimal values of displacements and optimized violations.
    
    \subsection{Two-Mode Squeezed Vacuum States}\label{A:Two-Mode Squeezed Vacuum State}
    
    We aim to maximize the violation of the BCCB inequality (\ref{genCHSH}) for such states with our MZI arrangement. As done previously for ECS in (\ref{E_ECS}), we calculate the expectation value of the observable $S$ (\ref{S_beta_gamma}) for $\ket{\Psi_{TMSV}(r)}$:
    
    \begin{widetext}
    \begin{align}\label{Psi_TMSV_S}
        \mathbb{E}(S)_{TMSV} = & \bra{\Psi_{TMSV}(r)} S\ket{\Psi_{TMSV}(r)} \nonumber \\ 
        = & \bra{\Psi_{TMSV}(r)}\Big(\sum_{i=1}^n A(\beta_i) \otimes A(\gamma_i) + \sum_{i=1}^{n-1} A(\beta_{i+1}) \otimes A(\gamma_i) - A(\beta_1) \otimes A(\gamma_n)\Big)\ket{\Psi_{TMSV}(r)}
    \end{align}
    Introducing $g(r,\beta_i,\gamma_j) = \bra{\Psi_{TMSV}(r)} A(\beta_i) \otimes A(\gamma_j)\ket{\Psi_{TMSV}(r)}$:
    \begin{align}\label{g_ij}
        g(r,\beta_i,\gamma_j) = 1-2\frac{e^{-|\beta_i|^2/\cosh^2{r}}+e^{-|\gamma_j|^2/\cosh^2{r}}-2\exp{-|\beta_i|^2-|\gamma_j|^2-2Re(\beta_i\gamma_j)\tanh{r}}}{\cosh^2{r}}
    \end{align}
    \end{widetext}

    We obtain the following:
    \begin{align} \label{E_TMSV}
        \mathbb{E}(S)_{TMSV} = & \sum_{i=1}^n g(r,\beta_i,\gamma_i) + \sum_{i=0}^{n-1} g(r,\beta_{i+1},\gamma_i) \nonumber \\
        - & g(r,\beta_{1},\gamma_n)
    \end{align}
    
    The numerical optimization is performed in the code \texttt{TMSV.py} and the results are stored in the file \texttt{max\_viol\_tmsv.pi}.  
    In the code \texttt{TMSV\_r.py}, we perform the optimization described as follows. For $n = 2, \dots, n$, we optimize the violation for fixed values of the squeezing parameter $r$. The results are stored in the file \texttt{tmsv\_r.pi}.
    
    The code \texttt{TMSV\_graphs.py} loads the data from \texttt{max\_viol\_tmsv.pi} and plots graphs of violation versus $r$ for $n = 2, \dots, 8$. Next, using the data from \texttt{max\_viol\_tmsv.pi} it produces graphs of maximal violation and the optimal value of parameter $r$ versus $n$.
    
    Finally, the code \texttt{Plots\_all.py} produces a plot comparing violations using data from \texttt{max\_viol\_eig.pi}, \texttt{max\_viol\_ecs.pi} and \texttt{max\_viol\_tmsv.pi}.
    
    
    

\end{document}